\RequirePackage{lineno}
\documentclass[aps,prd,twocolumn,superscriptaddress,amsmath,amssymb]{revtex4-2}

\usepackage{multirow}
 
\def\revtex{true}
\usepackage{etoolbox} % for checking definitions
\usepackage{graphicx} % Include figure files
\usepackage{epstopdf} % Allow to include eps files
\usepackage{dcolumn} % Align table columns on decimal point
\usepackage{xcolor} % Color support
\usepackage{amsmath,amssymb} % Extended symbol collection
\usepackage{hyperref} % Generate pdfs with hyperlinks
\usepackage[utf8]{inputenc} % Allow UTF8 characters
\usepackage[english]{babel} % All publications are in English
\usepackage{ifthen} % for conditional statements
\usepackage{blindtext} % For testing
\usepackage{orcidlink} % for ORCIDs in author list
\usepackage{datetime2} % for working version number
\ifthenelse{\isundefined{\revtex}}{\usepackage[displaymath, mathlines]{lineno}}{} % Line numbers for drafts

\graphicspath{{figures/}} % Use figures directory for figures

\newboolean{articletitles}
\setboolean{articletitles}{true} % False removes titles in references

\input{belle2-symbols}

\newcommand{\docnumber}{BELLE2-NOTE-XX-YYYY-ZZZ}  % Belle II document number
\newcommand{\analysis}{}  % Belle II analysis number
\newcommand{\pubnumber}{}  % Belle II publication number
\newcommand{\proponents}{}  % names of the proponents
\newcommand{\referees}{}  % names of the referees
\newcommand{\pcreader}{}  % name of the PC reader
\newcommand{\target}{}  % intended journal
\ifdefvoid{\draftdoc}{}{
    \ifdefvoid{\analysis}{
        
    }{

    }
}

\begin{document}

\ifdefvoid{\draftdoc}{}{\begin{widetext}\begin{flushright}
Document: \docnumber\\
Analysis: \ifdefvoid{\analysis}{}{\href{https://docs.belle2.org/pub_data/analyses/\analysis}{\analysis}}\\
Publication: \ifdefvoid{\pubnumber}{}{\href{https://authors.belle2.org/\#\pubnumber}{\#\pubnumber}}\\
Proponents: \proponents\\
Referees: \referees\\
PC Reader: \pcreader\\
Intended journal: \target\\
\end{flushright}
\vspace*{24pt}
\end{widetext}\linenumbers}

\title{\protect Measurements of $C\!P$ asymmetries in $D^+_{(s)} \to \eta h^+(h=\pi,K)$ and branching fractions of $D^+_{(s)}\to \eta K^+$ at Belle~II}

%%% Paper:    Measurements of $C\!P$ asymmetries in $D^+_{(s)} \to \eta h^+(h=\pi,K)$ and branching fractions of $D^+_{(s)}\to \eta K^+$ at Belle II
%%% Journal:  PhysRev
%%% Contacts: J.-Y. Kim, Y.-J. Kwon
%%% ====================================================================
%%% Use \input{pub125-orcid} to insert this material into your latex file.
  \author{M.~Abumusabh\,\orcidlink{0009-0004-1031-5425}} % 26883
  \author{I.~Adachi\,\orcidlink{0000-0003-2287-0173}} % 2590
% \author{K.~Adamczyk\,\orcidlink{0000-0001-6208-0876}} % 2239
  \author{A.~Aggarwal\,\orcidlink{0000-0002-5623-3896}} % 24463
% \author{L.~Aggarwal\,\orcidlink{0000-0002-0909-7537}} % 10083
  \author{H.~Ahmed\,\orcidlink{0000-0003-3976-7498}} % 11323
% \author{J.~K.~Ahn\,\orcidlink{0000-0002-5795-2243}} % 7423
  \author{Y.~Ahn\,\orcidlink{0000-0001-6820-0576}} % 14363
  \author{H.~Aihara\,\orcidlink{0000-0002-1907-5964}} % 2223
  \author{M.~Akdag\,\orcidlink{0009-0004-3728-1077}} % 27563
  \author{N.~Akopov\,\orcidlink{0000-0002-4425-2096}} % 9443
  \author{A.~Akram\,\orcidlink{0000-0003-0198-5852}} % 28483
  \author{S.~Alghamdi\,\orcidlink{0000-0001-7609-112X}} % 27804
% ...
% 13403
% \author{S.~Zou\,\orcidlink{0000-0003-3377-7222}} % 19363
\collaboration{Belle II Collaboration}

\begin{abstract}
% \blindtext
We report measurements of the $C\!P$ asymmetries of $D^+_{(s)} \to \eta \pi^+$ and $D^+_{(s)} \to \eta K^+$ decays, as well as the branching fractions of $D^+_{(s)} \to \eta K^+$.
The analysis uses $e^{+}e^{-}$ collision data collected with the Belle II detector corresponding to an integrated luminosity of $428\,\mathrm{fb}^{-1}$. 
The measured $C\!P$ asymmetries are
\begin{align*}
A_{C\!P}(D^+ \to \eta\pi^+)     &= (-0.70 \pm 5.15 \pm 0.18)\times 10^{-3}, \\
A_{C\!P}(D_s^+ \to \eta\pi^+)   &= (-3.23 \pm 3.80 \pm 0.18)\times 10^{-3}, \\
A_{C\!P}(D^+ \to \eta K^+)      &= (-1.63 \pm 6.51 \pm 0.26)\times 10^{-2}, \\
A_{C\!P}(D_s^+ \to \eta K^+)    &= (-0.37 \pm 1.85 \pm 0.10)\times 10^{-2}.
\end{align*}
The branching-fraction ratios of $D^+_{(s)} \to \eta K^+$ relative to $D^+_{(s)} \to \eta \pi^+$ are measured to be $\mathcal{B}(D^{+} \to \eta K^+)/\mathcal{B}(D^{+} \to \eta \pi^+)=(2.90 \pm 0.21 \pm 0.05)\times 10^{-2}$ and $\mathcal{B}(D_s^{+} \to \eta K^+)/\mathcal{B}(D_s^{+} \to \eta \pi^+)=(8.91 \pm 0.19 \pm 0.14)\times 10^{-2}$.
The first and second uncertainties are statistical and systematic, respectively.
\end{abstract}

\maketitle

% \linenumbers

\section{Introduction}

The study of charge-parity ($C\!P$) violation in charmed meson decays provides a sensitive test of the Standard Model (SM).
While $C\!P$ violation is well established in the strange and beauty meson sectors, SM asymmetries in charm decays are typically expected to be at the level of $\mathcal{O}(10^{-3})$ or below~\cite{Buccella1995,Golden:1989qx}.
The LHCb Collaboration reported the first observation of $C\!P$ violation in charm decays through a difference between the decay-rate $C\!P$ asymmetries of $D^0\to K^+K^-$ and $D^0\to \pi^+\pi^-$~\cite{LHCb:2019hro}. 
This observation motivates further measurements of $C\!P$ asymmetries in charmed meson decays.

The $D^+_{(s)}\to\eta h^+$ decays, where $h$ denotes $\pi$ or $K$, offer complementary probes of direct $C\!P$ violation.
(Charge-conjugate modes are implied throughout this paper, unless otherwise stated.)
Among these decays, $D_s^+ \to \eta \pi^+$ and $D^+ \to \eta K^+$ proceed via Cabibbo-favored and doubly Cabibbo-suppressed amplitudes, respectively.
The singly Cabibbo-suppressed decays $D^+ \to \eta \pi^+$ and $D_s^+ \to \eta K^+$ are of particular interest, as they receive contributions from both the $c \to d\bar{d}u$ and $c \to s\bar{s}u$ transitions~\cite{Cheng:2019ggx}, governed respectively by the Cabibbo-Kobayashi-Maskawa (CKM) factors $\lambda_d=V_{cd}^{*}V_{ud}$ and $\lambda_s=V_{cs}^{*}V_{us}$~\cite{Kobayashi:1973}.
The relative weak phase between these CKM factors is small in the SM, with $\mathrm{Im} (\lambda_d / \lambda_s ) \sim \mathcal{O}(10^{-4}\text{--}10^{-3})$.
This phase originates from the complex phase of the CKM matrix and appears at higher order in its Wolfenstein expansion~\cite{Wolfen:1983}.
Together with strong-phase differences from final-state interactions, this weak phase can generate nonzero direct $C\!P$ asymmetries~\cite{Cheng:2012xb,Li:2012cfa}.

We report measurements of the $C\!P$ asymmetries in $D_{(s)}^+\to \eta h^+$ decays.
In addition, we measure the branching fractions of $D_{(s)}^+\to \eta K^+$ decays relative to their corresponding $\eta \pi^+$ normalization modes.
This analysis uses $D^+$ and $D^+_s$ mesons produced in $e^+e^-$ collisions recorded by Belle II at or near the $\Upsilon(4S)$ resonance between 2019 and 2022, corresponding to an integrated luminosity of $428\,\mathrm{fb}^{-1}$.

The $C\!P$ asymmetry is defined as 
\begin{equation}
A_{C\!P}(D^+_{(s)} \to f^+) \equiv \frac{ \Gamma (D^+_{(s)} \to f^+) - \Gamma(D^-_{(s)} \to f^-) }{ \Gamma(D^+_{(s)} \to f^+) + \Gamma(D^-_{(s)} \to f^-) }, 
\end{equation}
where $\Gamma$ denotes the decay rate and $f$ represents the final state.
Previous measurements of $C\!P$ asymmetries in $D^+_{(s)} \to \eta h^+$ decays, as reported by LHCb~\cite{LHCb2023_CPAsym_DstoEtaPi,LHCb2021_CPViolation_DtoHPi0Eta}, BESIII~\cite{BESIII:2024oth}, Belle~\cite{Belle2021_CPAsym_DstoEtaPi,Belle2011_DtoEtaHp}, and CLEO~\cite{Cleo2010_DtoEtaHp}, are consistent with $C\!P$ symmetry.
The raw asymmetry is measured from the fitted yields of positively and negatively charged charmed mesons,
\begin{equation}
A_{\mathrm{raw}}^{f} \equiv \frac{N(D^+_{(s)} \to f^+) - N(D^-_{(s)} \to f^-)}{N(D^+_{(s)} \to f^+) + N(D^-_{(s)} \to f^-)},
\end{equation}
where $N$ is the fitted signal yield.

The raw asymmetry measured in data includes contributions from the physical $C\!P$ asymmetry, the forward-backward production asymmetry ($A_{\mathrm{FB}}$), and the detection asymmetry of the charged final-state hadron ($A_{\epsilon}$).
Assuming these asymmetries are small, it can be written as 
\begin{equation}
A_{\mathrm{raw}}^{\eta h^+} = A_{C\!P}(D^+_{(s)} \to \eta h^+) + A_{\mathrm{FB}}(D^+_{(s)}) + A_{\epsilon}(h^+).
\end{equation}
We use the Cabibbo-favored control modes $D^+ \to K_S^0\pi^+$ and $D_s^+ \to K_S^0K^+$ to correct for detection asymmetries, since no direct $C\!P$ violation is expected.
Their raw asymmetries are expressed as
\begin{equation}
A_{\mathrm{raw}}^{K_S^0 h^+} = A_{\mathrm{FB}}(D^+_{(s)}) + A_{\epsilon}(h^+) + A(\bar{K}^0),
\end{equation}
where $A(\bar{K}^0)$ accounts for the contribution from neutral-kaon mixing, including regeneration due to interactions with detector material~\cite{Pais:1955sm,Ko:2010mk}.

Charmed mesons in $e^+e^- \to c \bar{c}$ events are produced with a forward-backward asymmetry due to $\gamma -Z^0$ interference and higher-order effects~\cite{Berends:1973fd,Brown:1973ji,Cashmore:1985vp}.
This contribution is canceled by averaging the raw asymmetries measured in the forward and backward regions: 
\begin{equation}
A_{\mathrm{raw}}^{\prime} = \frac{ A_{\mathrm{raw}}(\cos\theta^{*} > 0) + A_{\mathrm{raw}}(\cos\theta^{*} < 0) }{2},
\label{eq:acp_raw_prime}
\end{equation}
 where $\theta^{*}$ is the polar angle of the $D^+_{(s)}$ momentum in the $e^+e^-$ center-of-mass (CM) frame.
The physical $C\!P$ asymmetry is then obtained from 
\begin{equation}
A_{C\!P}(D^+_{(s)} \to \eta h^+) = A_{\mathrm{raw}}^{\prime\eta h^+} - A_{\mathrm{raw}}^{\prime K_S^0 h^+} + A(\bar{K}^0).
\end{equation}

Alongside the $C\!P$ asymmetries, the branching fractions of $D^+_{(s)}\to \eta K^+$ are measured relative to the corresponding $D^+_{(s)}\to \eta \pi^+$ normalization modes in order to reduce common systematic uncertainties from luminosity, charm production, and reconstruction effects.
The branching fraction ratio is calculated as
\begin{equation}
\label{br_ratio_equation}
\frac{ \mathcal{B}(D^+_{(s)} \to \eta K^+) }{ \mathcal{B}(D^+_{(s)} \to \eta \pi^+) } = \frac{ N_{\mathrm{sig}}^{\eta K^+}/\epsilon^{\eta K^+} }{ N_{\mathrm{sig}}^{\eta \pi^+}/\epsilon^{\eta \pi^+} },  
\end{equation}
where $N_{\mathrm{sig}}$ is the fitted signal yield and $\epsilon$ is the total selection efficiency.
The branching fractions are determined over the entire $\cos\theta^{*}$ region.
The Belle and BESIII Collaborations have previously reported branching-fraction measurements for both $D^+ \to \eta K^+$~\cite{Belle2011_DtoEtaHp,Bes2018_DptoEtaKp,Bes2025_DptoEtaKp} and $D^+_s \to \eta K^+$~\cite{Belle2021_CPAsym_DstoEtaPi,Bes2020_DsptoEtaKp}.

The $\eta$ meson is reconstructed in both the $\eta \to \gamma \gamma\,(\eta_{\gamma\gamma})$ and $\eta \to \pi^+ \pi^- \pi^0\,(\eta_{3\pi})$ decay modes, with $\pi^0 \to \gamma \gamma$.
The results from the two $\eta$ decay modes are obtained independently and subsequently combined using the best linear unbiased estimator method, with correlated uncertainties taken into account~\cite{Lyons1988, Valassi2003}.

\section{Belle II detector and simulation}
The Belle II detector is located at the SuperKEKB asymmetric-energy $e^+e^-$ collider, which operates with CM energies at and near the $\Upsilon(4S)$ resonance~\cite{Akai:2018}.
The Belle II detector~\cite{Abe:2010} has a cylindrical geometry around the interaction point.
Closest to the beam pipe are a two-layer silicon-pixel detector~(PXD), a four-layer double-sided silicon-strip detector~(SVD)~\cite{Belle-IISVD:2022}, and a 56-layer central drift chamber~(CDC).
These tracking detectors provide position measurements and are used to reconstruct charged-particle trajectories~\cite{Bertacchi:2020eez}.
For the data sample used in this analysis, the second layer of the PXD provided 15\% coverage of the azimuthal angle.
The symmetry axis of the detector defines the $z$ axis and is approximately aligned with the electron-beam direction.
Surrounding the CDC is a time-of-propagation counter~(TOP)~\cite{Kotchetkov:2018qzw} in the central region and an aerogel-based ring-imaging Cherenkov counter~(ARICH) in the forward region.
They provide charged-particle identification in combination with the energy-loss information from the SVD and CDC.
Surrounding the TOP and ARICH is an electromagnetic calorimeter~(ECL) composed of CsI(Tl) crystals, which primarily provides energy and timing measurements for photons and electrons.
The ECL is enclosed by a superconducting solenoid that provides a 1.5~T magnetic field parallel to the $z$ axis.
The magnetic flux return is instrumented with resistive-plate chambers and plastic scintillator modules for the detection of muons, $K_L^0$ mesons, and neutrons.

We use simulated event samples to study background contributions, optimize the selection requirements, develop fitting models, and validate the analysis procedure.
Simulated continuum $e^+e^- \to q\bar{q}$ events, where $q=u,d,c,$ or $s$, are produced with \kkmc~\cite{Jadach:1999vf} and hadronized with \pythia~\cite{Sjostrand:2014zea}.
Simulated $e^+e^- \to \Upsilon(4S)$ events are generated, and subsequent particle decays are modeled using \evtgen~\cite{Lange:2001uf} interfaced to \pythia~\cite{Sjostrand:2014zea}.
The $e^+e^- \to \tau^+\tau^-$ samples are generated with \kkmc, and the subsequent $\tau$ decays are simulated by \tauola~\cite{Jadach:1990mz}.
Final-state radiation is modeled using \photos~\cite{Barberio:1990ms}, and the detector response is simulated with \geant~\cite{Agostinelli:2002hh}.
Beam backgrounds are included by overlaying simulated events with minimum-bias data recorded using a random trigger issued after a fixed delay following a Bhabha event~\cite{Abe:2010}.
Both data and simulated samples are reconstructed and analyzed with the Belle II Analysis Software Framework~\cite{Kuhr:2018lps,basf2-zenodo}.

\section{Event selection}
\label{sec:event_selection}

Candidate events are selected in two steps.
Loose preselection requirements are first applied to remove poorly reconstructed candidates and reduce large combinatorial backgrounds.
A boosted decision tree (BDT), implemented with XGBoost~\cite{Chen:2016btl}, is then used to further suppress the remaining background.
This background predominantly arises from incorrect combinations of final-state particles in continuum events, rather than from specific decay modes.

The same strategy is used for the signal modes and for the corresponding control and normalization modes.
For the $A_{C\!P}$ measurements, one BDT classifier is trained for each final state:
$D^+_{(s)}\to\eta_{\gamma\gamma}\pi^+$,
$D^+_{(s)}\to\eta_{3\pi}\pi^+$,
$D^+_{(s)}\to\eta_{\gamma\gamma}K^+$, and
$D^+_{(s)}\to\eta_{3\pi}K^+$.
The BDTs are trained using simulated candidates that satisfy the preselection.
Correctly reconstructed signal candidates constitute the signal sample, while the background sample contains all other candidates, including misreconstructed candidates from signal decays.
In the branching fraction measurements, the BDT classifier trained for the $D^+_{(s)}\to\eta K^+$ signal mode is also applied to the corresponding $D^+_{(s)}\to\eta\pi^+$ normalization mode, with the BDT requirement chosen to give the same selection efficiency for the two modes.

The offline reconstruction begins with a hadronic event selection, which rejects events consistent with Bhabha scattering and requires at least three charged tracks with transverse momentum greater than $0.2\,\mathrm{GeV}/c$ that are consistent with originating from the $e^+e^-$ interaction region.
The selected events are then used to reconstruct the individual decay modes described below.

\subsection{\texorpdfstring{$D^+_{(s)}\to \eta h^+(h=\pi,K)$}{D(s)+ to eta h+(h=pi,K)}}
\label{sec:event_selection_signal}
Unless otherwise stated, all kinematic quantities are evaluated in the laboratory frame.
Charged-particle candidates are required to be within the CDC acceptance $(17^{\circ} < \theta < 150^{\circ})$ and to have at least one associated CDC hit.
The distance of closest approach of the tracks to the $e^+e^-$ interaction point must be less than $3\,\mathrm{cm}$ along the beam direction and less than $1\,\mathrm{cm}$ in the transverse plane.
The primary pion (kaon) from the charmed meson decay, hereafter referred to as the hard pion (kaon), is required to have momentum greater than $0.4\,\mathrm{GeV}/c$.
Charged particles are identified as pions or kaons using information from all subdetector systems except the PXD~\cite{Belle-II:2025tpe}. 
The efficiencies are approximately 81\%, 60\%, and 99\% for the hard pion, hard kaon, and charged pions from the $\eta_{3\pi}$ decays, respectively; the corresponding pion--kaon misidentification probabilities are about 5\%, 2\%, and 40\%.
The tight kaon-identification selection is used to suppress pion-to-kaon misidentification, resulting in a lower efficiency.
For the charged pions from the $\eta_{3\pi}$ decay, a looser identification selection is used because the subsequent $\eta$ invariant-mass selection provides sufficient suppression of combinatorial background.

Photon candidates are reconstructed from ECL clusters that are not matched to charged particles, lie within the CDC angular acceptance, and satisfy requirements on energy and cluster quality.
Each photon candidate must have energy deposits in at least two ECL crystals and an energy exceeding $55\,\mathrm{MeV}$ for $\pi^0$ reconstruction or $100\,\mathrm{MeV}$ for $\eta_{\gamma\gamma}$ reconstruction.
To suppress photon clusters caused by beam background or hadronic showers, two multivariate classifiers~\cite{Cheema:2024iek} are utilized.
The neutral-pion candidates are reconstructed from photon pairs and are required to have an invariant mass within $[0.12, 0.145]\,\mathrm{GeV}/c^2$ and a converged kinematic fit. The two photons forming the candidate must be separated by less than $1.5\,\mathrm{rad}$ in azimuth, with the opening angle between the two photon momenta smaller than $1.4\,\mathrm{rad}$.
The $\eta_{\gamma\gamma}$ candidates are required to have an invariant mass within $[0.52, 0.57]\,\mathrm{GeV}/c^2$ and a momentum greater than $0.4\,\mathrm{GeV}/c$.
The two photons forming the candidate must be separated by less than $1.8\,\mathrm{rad}$ in azimuth, with the opening angle smaller than $1.6\,\mathrm{rad}$.
To suppress backgrounds from $\pi^0$ decays, each photon used in the $\eta_{\gamma\gamma}$ reconstruction is combined with any photon not assigned to the signal $\eta_{\gamma \gamma}$, and the candidate is retained only if each resulting invariant mass satisfies $|M(\gamma\gamma)-m_{\pi^0}| > 11\,\mathrm{MeV}/c^2$, where $m_{\pi^0}$ denotes the nominal mass of $\pi^0$~\cite{PDG:2026}.
This veto retains approximately 83\% (87\%) of $D^+$ ($D^+_s$) signal candidates, with negligible dependence on whether the final-state hard meson is $\pi^+$ or $K^+$.
It rejects about 55\% (57\%) of the total background for the $\eta \pi^+$ ($\eta K^+$) modes.
The $\eta_{3\pi}$ candidates are required to have an invariant mass within $[0.535, 0.57]\,\mathrm{GeV}/c^2$ and a momentum greater than $0.4\,\mathrm{GeV}/c$.

The $D^+_{(s)}$ candidates are formed by combining an $\eta$ candidate with a hard $\pi^+$ or $K^+$, and their invariant mass is required to be in the range $[1.6, 2.1]\,\mathrm{GeV}/c^2$.
A kinematic fit constrains the final-state particles to a common decay vertex, constrains the $\pi^0$ and $\eta$ candidate masses to their nominal values, and constrains the charmed meson candidate to originate from the interaction point~\cite{Krohn:2019dlq}.
Only candidates with a $\chi^2$ probability greater than 0.001 are retained.
To effectively suppress combinatorial backgrounds and contributions from charmed mesons produced in $B$ meson decays, the CM-frame momentum of the $D^+_{(s)}$ candidate is required to be greater than $2.5\,\mathrm{GeV}/c$.

The BDT input variables are based on the vertexing and kinematic properties of the charmed meson candidates.
They include the transverse impact parameter of the hard track with respect to the interaction point, the impact parameter of the charmed meson along the beam direction, and the cosine of the angle between the charmed meson momentum and the vector from the interaction point to its decay vertex, projected onto the plane perpendicular to the beam axis.
The cosine of the helicity angle of the hard $\pi^+$ or $K^+$ and the CM-frame momentum of the charmed meson are also used.
The helicity angle is defined using the momentum of the hard $\pi^+$ or $K^+$ in the charmed meson rest frame with respect to the direction opposite to the boost from the laboratory frame.
An additional variable, the photon-energy asymmetry, is used only for the $\eta_{\gamma\gamma}$ modes.
When the corresponding selection is applied to the $K_S^0h^+$ control channels, this variable is replaced by the momentum asymmetry of the two pion tracks from the $K_S^0$ candidate.

The BDT selection requirement is optimized by scanning the selection threshold and maximizing $N_{\rm sig}/\sigma_{N_{\rm sig}}$, where $N_{\rm sig}$ is the fitted signal yield and $\sigma_{N_{\rm sig}}$ is its statistical uncertainty.
The requirement for the $\eta\pi^+$ modes is optimized using the $D^+$ signal, which is the singly Cabibbo-suppressed decay mode of primary interest.
In contrast, a balanced selection is used for the $\eta K^+$ modes, with the threshold set to the midpoint between the values separately optimized for the $D^+$ and $D_s^+$ signals.
Furthermore, the simulated samples of the $D_s^+$ modes are corrected using the CM-frame momentum-dependent reweighting factors described in Sec.~\ref{sec:reweight_MC}.

A small fraction of events contain multiple $D^+_{(s)}$ candidates after BDT selection, mainly due to incorrect photon combinations in the $\eta$ reconstruction.
The fraction ranges from about 1\% to 5\%, with larger values for the $\eta_{3\pi}$ modes than for the $\eta_{\gamma\gamma}$ modes.
In such events, a single candidate is retained by choosing the candidate with the highest kinematic-fit $\chi^2$ probability.
According to simulation, this best-candidate selection chooses the correct signal candidate in about 61\%--76\% of multiple-candidate events, depending on the final state.
The lower end of this range corresponds mainly to the $\eta_{\gamma\gamma}$ modes, while the higher end corresponds mainly to the $\eta_{3\pi}$ modes.

\subsection{\texorpdfstring{$D^+\to K_S^0\pi^+$ and $D_s^+\to K_S^0K^+$}{D+ to Ks pi and Ds to Ks K}}

The control decays $D^+ \to K_S^0 \pi^+$ and $D_s^+ \to K_S^0 K^+$ are reconstructed by combining a $K_S^0$ candidate with a charged pion or kaon candidate.
The $K_S^0$ candidates are formed from pairs of oppositely charged tracks, which are assigned the pion mass and constrained to originate from a common vertex.
The invariant mass of the pion pair is required to satisfy $|M(\pi^+\pi^-)-m_{K_S^0}| < 15\,\mathrm{MeV}/c^2 $, where $m_{K_S^0}$ is the nominal mass of $K_S^0$~\cite{PDG:2026}.
The ratio of its flight length to the corresponding uncertainty of the $K_S^0$ candidate is required to be larger than 10, and its momentum is required to be greater than $0.4\,\mathrm{GeV}/c$.

The $K_S^0 h^+$ candidates are subjected to the same kinematic and vertex fits used for the signal modes without any mass constraints.
The selections for the control modes are chosen to be as close as possible to those for the corresponding $D^+_{(s)}\to\eta h^+$ signal modes.
In particular, the charged pion and kaon candidates from the charmed meson decays are required to satisfy the same tracking, momentum, and particle-identification criteria as those used in the signal reconstruction.
This common treatment reduces possible differences in the detection asymmetry of the charged hadron.

\subsection{Validation of the BDT selection}
\label{sec:reweight_MC}

To estimate the BDT output difference between data and simulation, the control channels are studied before applying the BDT selection requirement.
Signal-enriched distributions are obtained from invariant-mass fits using the sPlot method~\cite{Pivk:2004ty}, with the charmed meson invariant mass used as the discriminating variable.
The BDT selection efficiency for \(D^+\to K_S^0\pi^+\) is found to be in good agreement between data and simulation.
By contrast, a residual difference is observed in the CM-frame momentum of \(D_s^+\to K_S^0K^+\) candidates.
Since the CM-frame momentum is one of the BDT input variables, this difference is found to affect the BDT output.
The simulated \(D_s^+\to K_S^0K^+\) sample is therefore reweighted as a function of the CM-frame momentum, using weights derived from an independent, high-purity, and well-populated sample of \(D_s^+\to\phi(\to K^+ K^-)\pi^+\) decays.
The weights are obtained from the ratio of the normalized sWeighted CM-frame distributions in data and simulation.
After applying this correction, the agreement between data and simulation in the BDT output distribution improves.
The data-to-simulation ratio of the BDT selection efficiency changes from \(1.101\pm0.004\) to \(1.003\pm0.004\) for \(\eta_{\gamma\gamma}K^+\), and from \(1.070\pm0.003\) to \(1.000\pm0.003\) for \(\eta_{3\pi}K^+\).
The same reweighting factors are applied to the relevant $D^+_s$ simulated samples used for efficiency evaluations, simulation-based validation, and systematic studies.

\section{Asymmetry determination}
The raw asymmetries are determined separately in the two regions $\cos\theta^*>0$ and $\cos\theta^*<0$, corresponding to forward and backward production of the charmed meson candidate in the $e^+e^-$ CM frame.
All fits described in this section are validated using ensembles of pseudoexperiments generated from the corresponding nominal fit models.
No significant fit bias is observed, and the statistical uncertainties are found to provide appropriate statistical coverage.

\subsection{\texorpdfstring{$D^+_{(s)}\to \eta h^+(h=\pi,K)$}{D(s)+ to eta h+(h=pi,K)}}

The raw asymmetries for $D^+_{(s)}\to\eta h^+$ are extracted from maximum-likelihood fits to the invariant-mass distributions of $\eta h^\pm$ candidates.
We apply binned fits to the $\eta\pi^+$ channels to save computational time, while using unbinned fits for the $\eta K^+$ channels.
The fit ranges for the $\eta\pi^+$ and $\eta K^+$ channels are $[1.71, 2.06]\,\mathrm{GeV}/c^2$ and $[1.75, 2.045]\,\mathrm{GeV}/c^2$, respectively.

The $M(\eta\pi^+)$ fit model consists of the $D^+$ and $D^+_s$ signals, a peaking background from $D_s^+\to\rho^+\eta$ with $\rho^+\to\pi^+\pi^0$, and combinatorial background.
The peaking background appears below the $D^+$ mass because the $\pi^0$ is not reconstructed.
To fit the $M(\eta K^+)$ distributions, the fit model incorporates $D^+$ and $D^+_s$ signals, a component from $D^+\to \eta \pi^+$ decays in which the pion is misidentified as a kaon, and combinatorial background.
Each fit component is described by a probability density function (PDF) whose shape parameters are shared between the two charge samples unless otherwise specified, an assumption tested in the systematic study.

The signal PDFs for the $D^+$ and $D_s^+$ components are modeled with a double-sided Crystal Ball function~\cite{Gaiser:Phd,Skwarnicki:1986xj} convolved with a Gaussian resolution function.
The means of the Gaussian resolution functions are fixed to zero, while the means of the double-sided Crystal Ball functions are allowed to float.
The Gaussian widths for the $D^+$ and $D_s^+$ signals are allowed to float independently in the $M(\eta\pi^+)$ fits.
Those in the $M(\eta K^+)$ fits are fixed to values obtained from simulation, while a common scale factor is allowed to float to account for residual resolution differences.
The remaining signal-shape parameters are determined from simulated signal samples after the full event selection.

The combinatorial background in the $M(\eta\pi^+)$ fits is modeled with an exponential function.
The $D_s^+\to\rho^+\eta$ peaking background is modeled with a Novosibirsk function~\cite{Belle:1999bhb}.
Its shape parameters are fixed from simulation, except for the peak position, which is allowed to float.
The fraction of the total background assigned to the $D_s^+\to\rho^+\eta$ component is also floated.

A second-order Chebyshev polynomial is used to model the combinatorial background in the $M(\eta K^+)$ fits.
The linear Chebyshev coefficient is fitted separately for the two charge samples, as ensemble tests show that this improves fit stability.
The misidentified $D^+\to\eta\pi^+$ background, in which the pion is reconstructed as a kaon, is modeled using a nonparametric kernel-density estimator~\cite{Cranmer:2000du} constructed from simulated samples.
Its yield is fixed to the simulation expectation after applying the scale factors determined from the corresponding normalization modes.
The scale factors are found to be $1.30\pm 0.01$ and $1.22\pm 0.01$ for $D^+ \to \eta_{\gamma\gamma} \pi^+$ and $D^+ \to \eta_{3\pi} \pi^+$, respectively.
The raw asymmetry of this misidentified component is fixed to zero, as ensemble tests show that the limited yield of this component does not provide sufficient sensitivity to determine its asymmetry.
The associated systematic uncertainty is evaluated separately.

\begin{figure*}[t]
  \centering
  \begin{tabular}{cc}
    \includegraphics[width=0.48\textwidth]{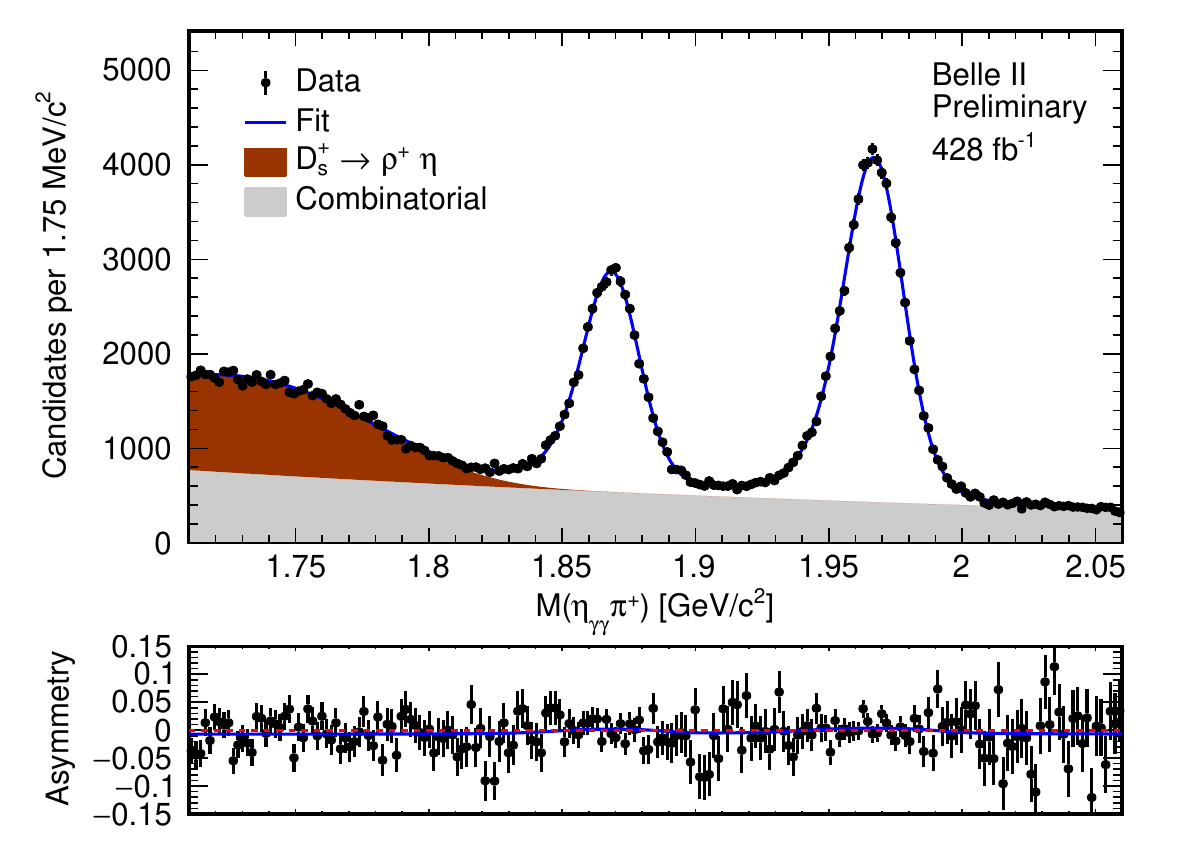}
    &
    \includegraphics[width=0.48\textwidth]{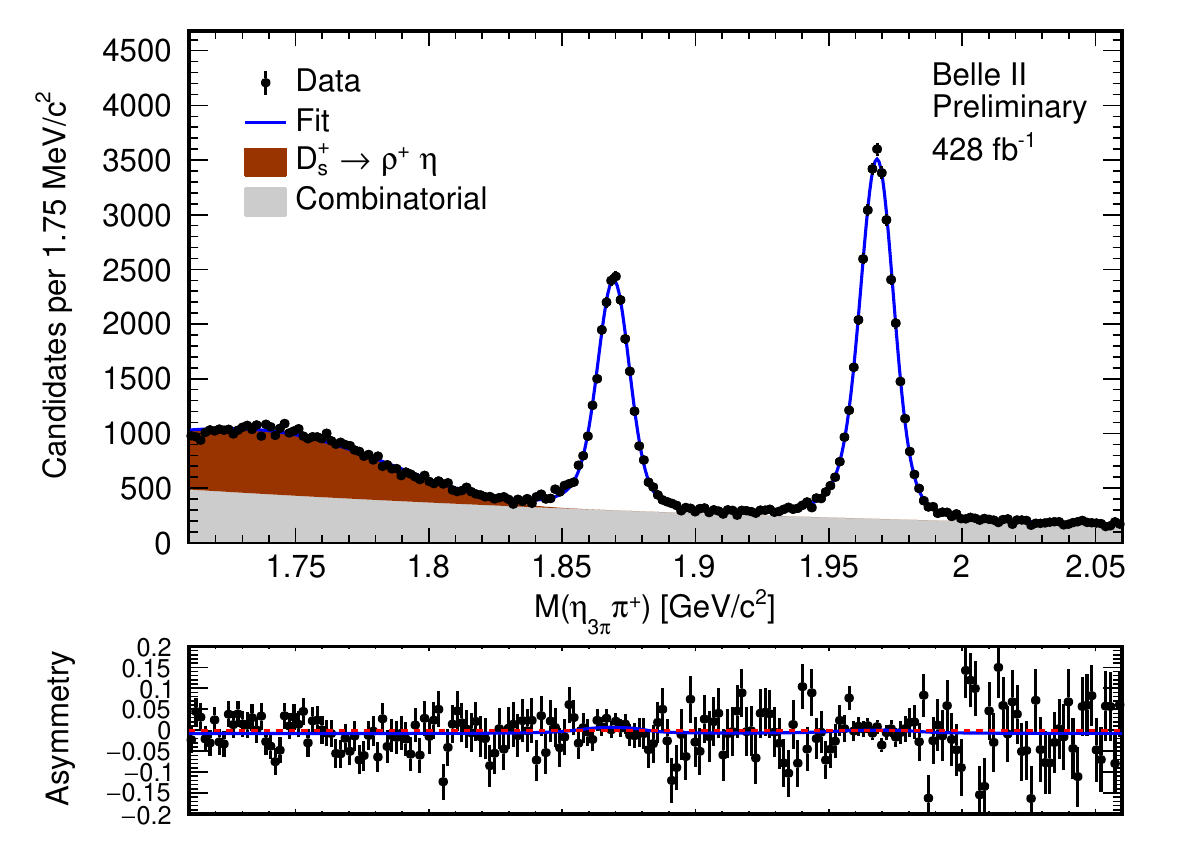}
    \\
    \includegraphics[width=0.48\textwidth]{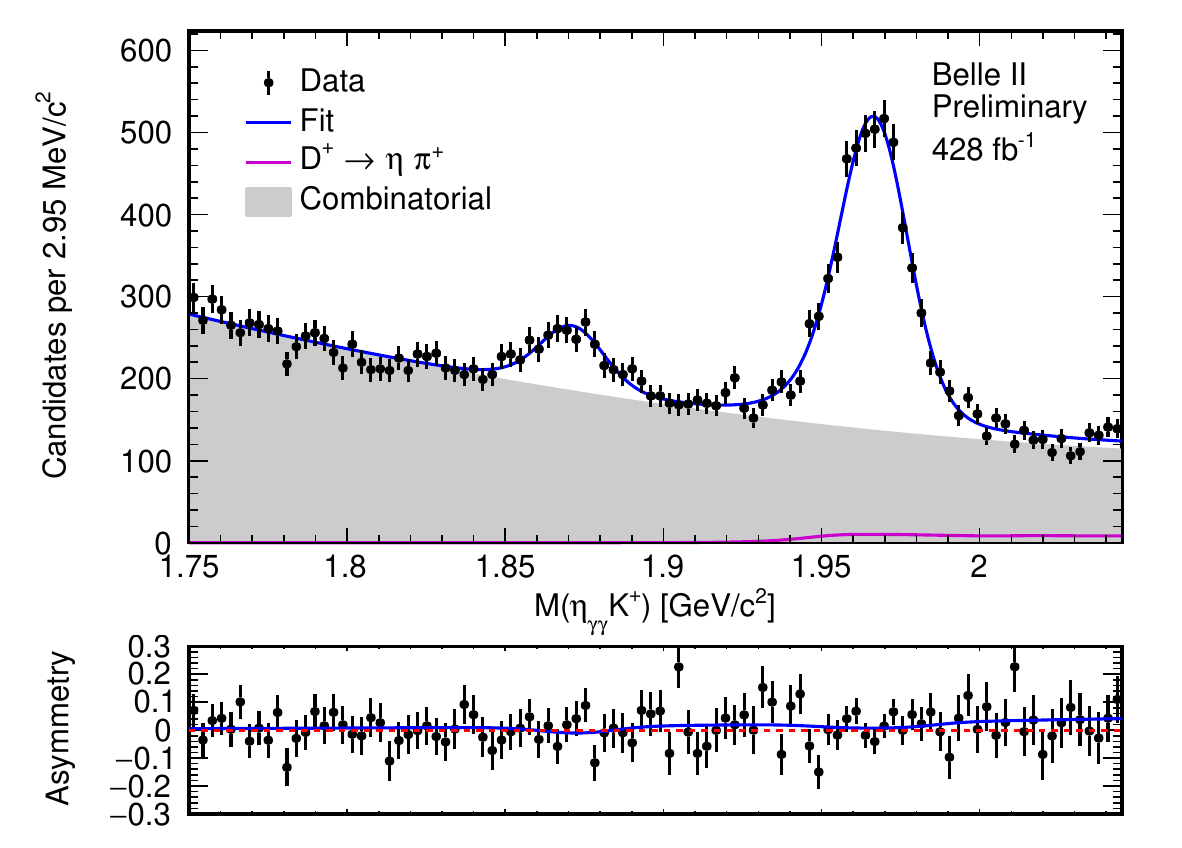}
    &
    \includegraphics[width=0.48\textwidth]{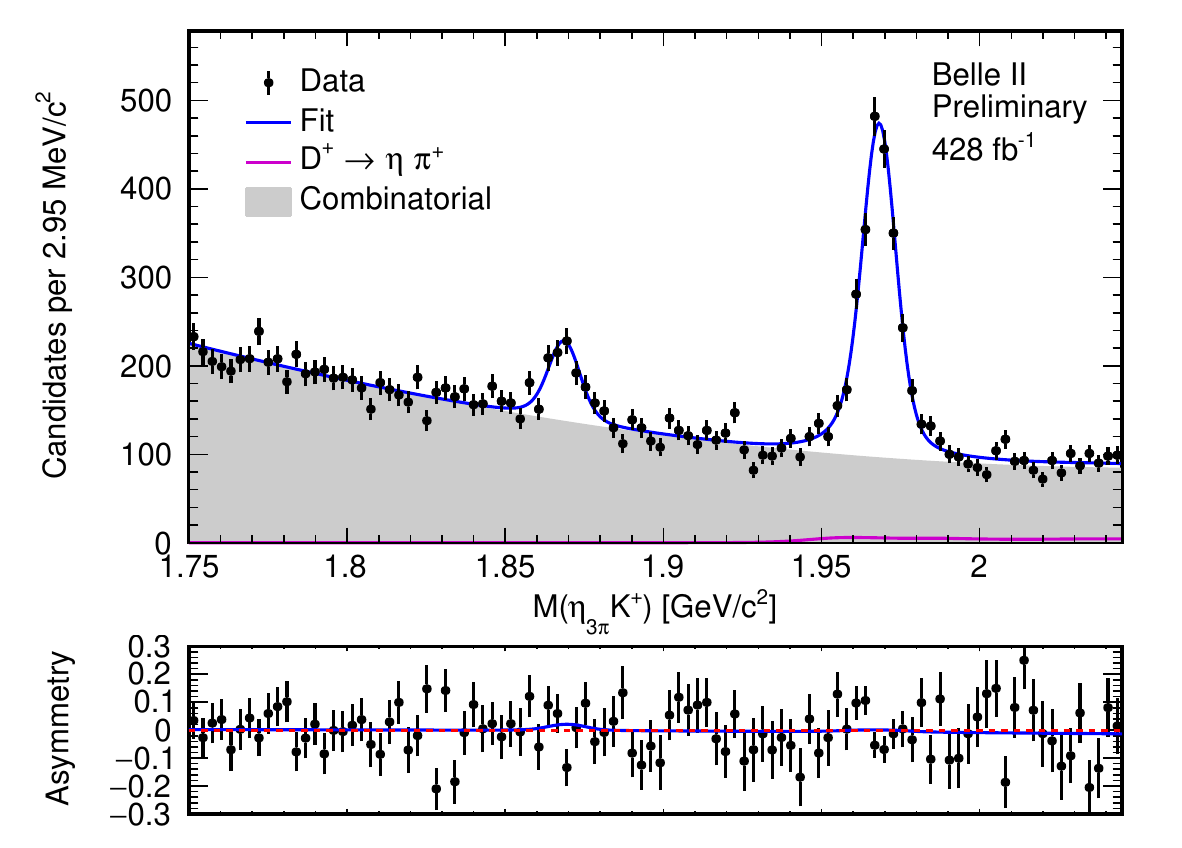}
  \end{tabular}
  \caption{
    Invariant mass distributions of the signal samples used for the $A_{C\!P}$ measurements. 
    The upper and lower rows show $D^+_{(s)} \to \eta\pi^+$ and $D^+_{(s)} \to \eta K^+$ candidates, with the left and right columns corresponding to $\eta_{\gamma\gamma}$ and $\eta_{3\pi}$, respectively.
    The points show the sums of the forward and backward samples and of the charge-conjugate modes.
    The curves show the corresponding sums of the fit projections obtained from independent fits in the two $\cos\theta^*$ regions.
    The lower panel shows the raw asymmetry as a function of mass, representing the average of the two fitted asymmetry curves and corresponding to the quantity defined in Eq.~\ref{eq:acp_raw_prime}.
    The dashed red line indicates zero asymmetry.
  }
  \label{fig:signal_fits}
\end{figure*}

The yields of positively and negatively charged candidates for each signal component $X=D,D_s$ are parameterized as
$N_X^{q} = \frac{1}{2}N_X\left(1+qA_{X,\rm raw}\right)$,
where $q=+1$ ($-1$) corresponds to positively (negatively) charged candidates.
The floated parameters include the total signal yields, the raw signal asymmetries, the combinatorial-background yield and asymmetry, and the relevant background-shape parameters.
The peaking-background fraction is additionally floated in the $M(\eta\pi^+)$ fits.
Table~\ref{tab:Aprime_data_all} lists the fitted raw asymmetries, and Fig.~\ref{fig:signal_fits} shows the corresponding invariant-mass distributions with fit projections.

\subsection{\texorpdfstring{$D^+\to K_S^0\pi^+$ and $D_s^+\to K_S^0K^+$}{D+ to Ks pi and Ds to Ks K}}

The raw asymmetries of the control channels are obtained with simultaneous unbinned maximum-likelihood fits to the invariant-mass distributions of the positively and negatively charged samples.
The fit ranges for the $K_S^0\pi^+$ and $K_S^0K^+$ final states are $[1.80, 1.93]\,\mathrm{GeV}/c^2$ and $[1.90, 2.03]\,\mathrm{GeV}/c^2$, respectively.
The $D^+\to K_S^0\pi^+$ sample uses the BDTs trained for $\eta\pi^+$ modes, and the $D_s^+\to K_S^0K^+$ sample uses those trained for the $\eta K^+$ modes.

The fit model contains a signal component and a combinatorial-background component.
The signal PDF is described by a Johnson $S_U$ function~\cite{Johnson:1949zj} convolved with a Gaussian resolution function, while the combinatorial background is modeled with an exponential function.
The two charge samples share all shape parameters, with the Gaussian mean fixed to zero.
As in the signal-mode fits, the signal and combinatorial-background yields are parameterized separately for the two candidate charges using component-dependent raw asymmetries.
The free parameters are the component yields and raw asymmetries, together with the remaining shape parameters.

The correction from the control channels also includes the neutral-kaon-induced asymmetry, denoted by $A({\bar{K}^0})$.
This asymmetry arises from $K^0$-$\bar{K}^0$ mixing, the different nuclear interaction cross sections of $K^0$ and $\bar{K}^0$ with detector material, and the interference between these effects.

The $A({\bar{K}^0})$ value is estimated as a function of the $K_S^0$ flight length, momentum, and azimuthal angle, following the procedure described in Ref.~\cite{LHCb:2014kcb} and also employed in Ref.~\cite{Belle-II:2025wsy}.
The expected asymmetry is evaluated using the Belle~II detector material distribution model.
To suppress combinatorial-background contamination, the evaluation is performed in signal-dominated mass regions.
The invariant-mass windows used for the $K_S^0\pi^+$ and $K_S^0K^+$ candidates are $[1.850, 1.885]\,\mathrm{GeV}/c^2$ and\ $[1.950, 1.985]\,\mathrm{GeV}/c^2$, respectively.
The background contributions in these regions are found to be negligible.

\begin{figure*}[t]
  \centering
  \begin{tabular}{cc}
    \includegraphics[width=0.48\textwidth]{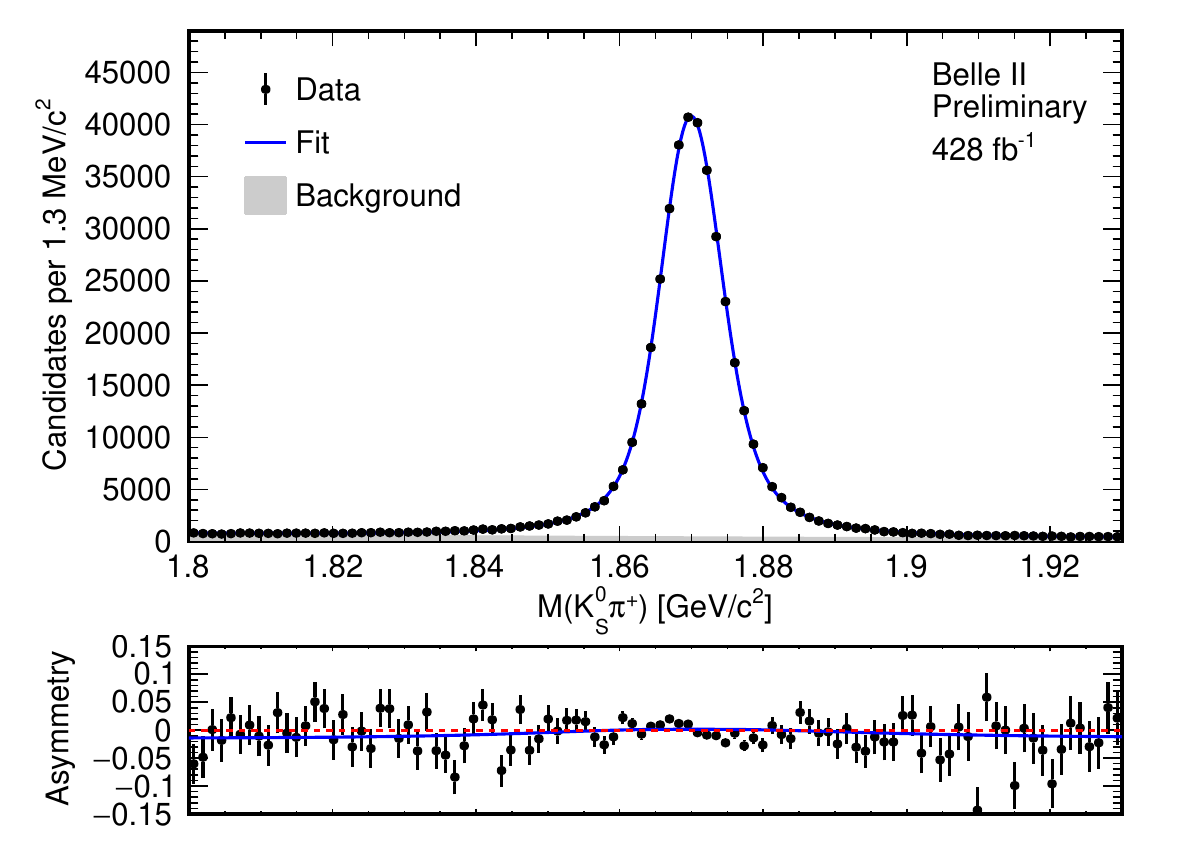}
    &
    \includegraphics[width=0.48\textwidth]{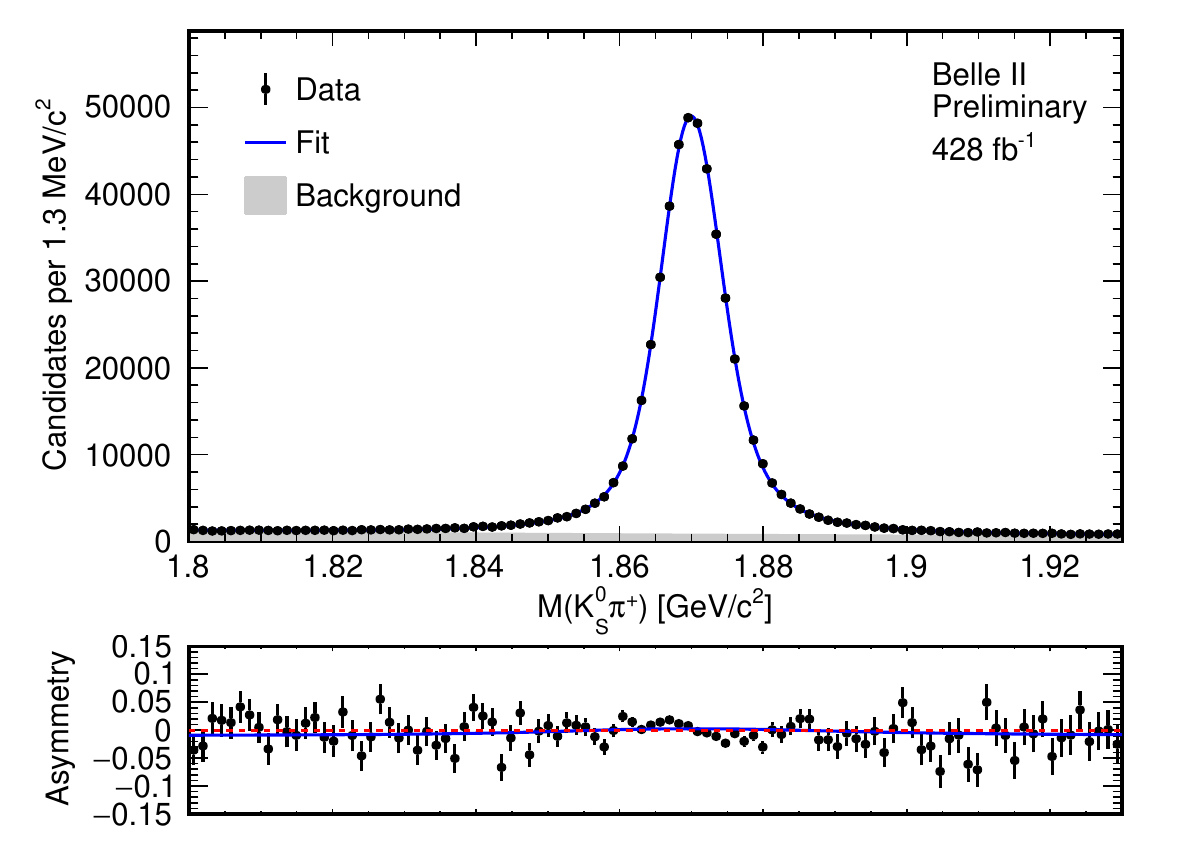}
    \\
    \includegraphics[width=0.48\textwidth]{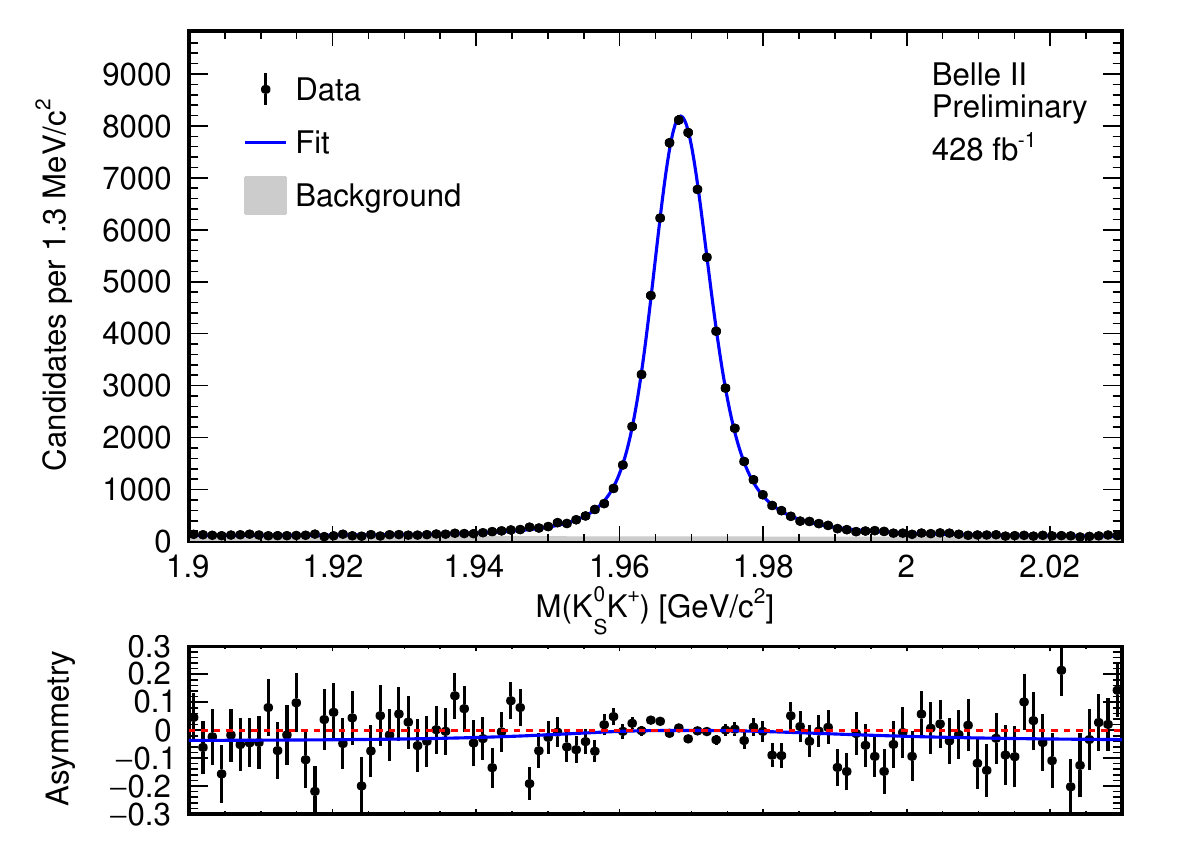}
    &
    \includegraphics[width=0.48\textwidth]{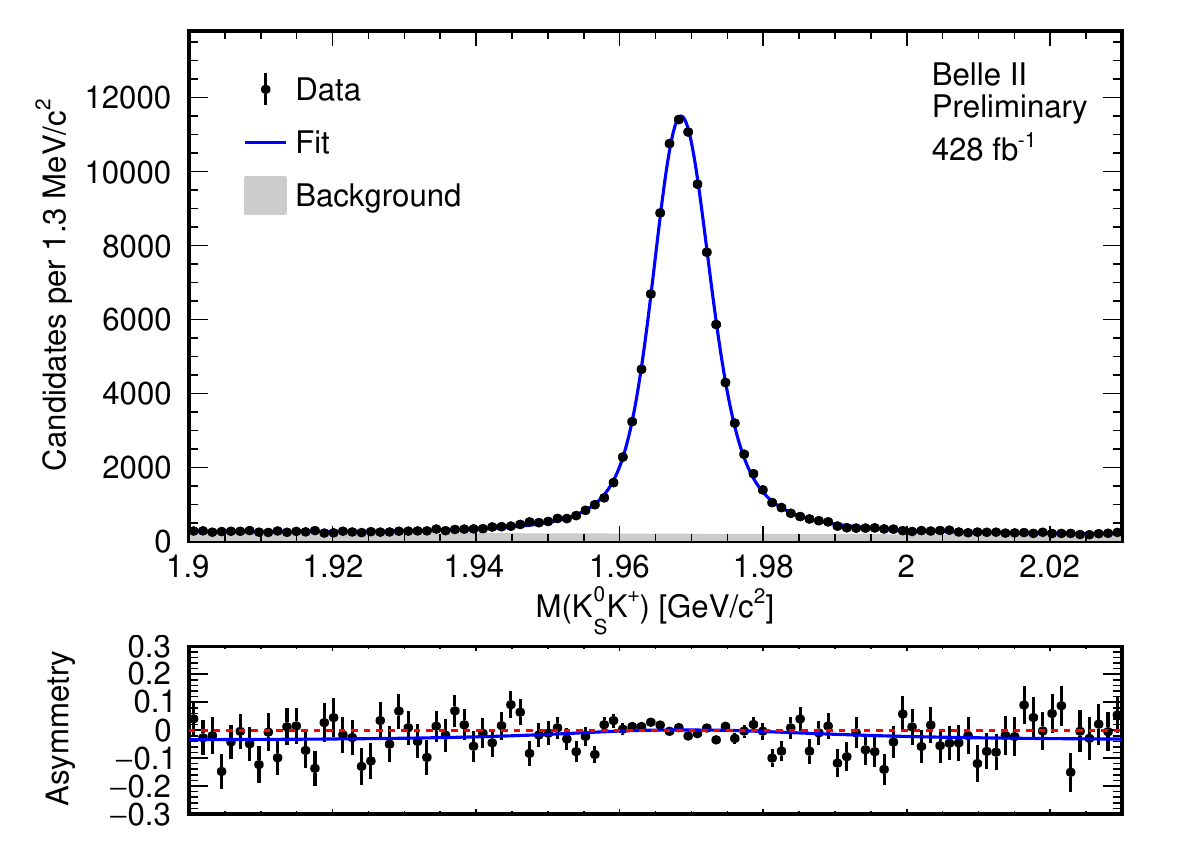}
  \end{tabular}
  \caption{
    Invariant mass distributions of the control samples used for the $A_{C\!P}$ measurements.
    The upper and lower rows show $D^+ \to K_S^0 \pi^+$ and $D^+_s\to K_S^0 K^+$ candidates, while the left and right columns use selections optimized for the $\eta_{\gamma\gamma}$ and $\eta_{3\pi}$ modes, respectively.
    The points show the sums of the forward and backward samples and of the charge-conjugate modes.
    The curves show the corresponding sums of the fit projections obtained from independent fits in the two $\cos\theta^*$ regions.
    The lower panel in each plot shows the raw asymmetry as a function of mass, with the solid curve representing the average of the two fitted asymmetry curves and corresponding to the quantity defined in Eq.~\ref{eq:acp_raw_prime}.
    The dashed red line indicates zero asymmetry.
  }
  \label{fig:ks_control_fit}
\end{figure*}

The evaluation is performed separately for each BDT configuration used in the corresponding signal analysis, so that the kinematic and detector conditions of the control samples match those of the signal samples as closely as possible.
The resulting values of $A({\bar{K}^0})$ for the $D^+\to K_S^0\pi^+$ control sample are $(-0.423 \pm 0.001)\times 10^{-2}$ and $(-0.421 \pm 0.001)\times 10^{-2}$ for the BDT selections associated with $\eta_{\gamma\gamma}\pi^+$ and $\eta_{3\pi}\pi^+$, respectively.
The corresponding values for the $D_s^+\to K_S^0K^+$ control sample are $(-0.428 \pm 0.003)\times 10^{-2}$ and $(-0.424 \pm 0.002)\times 10^{-2}$ for the BDT selections associated with $\eta_{\gamma\gamma}K^+$ and $\eta_{3\pi}K^+$, respectively.
Table~\ref{tab:Aprime_data_all} summarizes the fitted raw asymmetries for the control channels, and Fig.~\ref{fig:ks_control_fit} presents the corresponding invariant-mass distributions with fit projections.

\begin{table*}[t]
  \centering
  \caption{
    Values of $A^{(\prime)}_{\rm raw}$[$10^{-2}$] in data for the signal modes and the corresponding
    $A_{C\!P}$ control modes. All values are given in percent.
    The control modes are evaluated after applying the same BDT requirement as the
    corresponding signal mode. Therefore, the raw asymmetries of the same control
    channel can differ between the $\eta_{\gamma\gamma}$ and $\eta_{3\pi}$ samples.
  }
  \label{tab:Aprime_data_all}
  \begin{ruledtabular}
  \begin{tabular}{llccc}
    BDT mode & Decay mode
    & $A_{\rm raw}(\cos\theta^* > 0)$
    & $A_{\rm raw}(\cos\theta^* < 0)$
    & $A^{\prime}_{\rm raw}$ \\
    \hline
    \multirow{3}{*}{$\eta_{\gamma\gamma}\pi^+$}
      & $D^+ \to \eta_{\gamma\gamma}\pi^+$
      & $-1.218 \pm 0.833$ & $2.006 \pm 0.952$ & $0.394 \pm 0.633$ \\
      & $D_s^+ \to \eta_{\gamma\gamma}\pi^+$
      & $-1.537 \pm 0.595$ & $2.476 \pm 0.676$ & $0.470 \pm 0.450$ \\
      & $D^+ \to K_S^0\pi^+$
      & $-1.824 \pm 0.222$ & $2.178 \pm 0.250$ & $0.177 \pm 0.167$ \\
    \hline
    \multirow{3}{*}{$\eta_{3\pi}\pi^+$}
      & $D^+ \to \eta_{3\pi}\pi^+$
      & $-2.140 \pm 1.097$ & $3.799 \pm 1.219$ & $0.829 \pm 0.820$ \\
      & $D_s^+ \to \eta_{3\pi}\pi^+$
      & $-1.640 \pm 0.808$ & $1.707 \pm 0.891$ & $0.033 \pm 0.601$ \\
      & $D^+ \to K_S^0\pi^+$
      & $-1.823 \pm 0.205$ & $2.338 \pm 0.231$ & $0.258 \pm 0.155$ \\
    % \hline
    % \multicolumn{5}{c}{$D^+_{(s)} \to \eta K^+$ modes} \\
    \hline
    \multirow{3}{*}{$\eta_{\gamma\gamma}K^+$}
      & $D^+ \to \eta_{\gamma\gamma}K^+$
      & $-14.90 \pm 14.16$ & $-0.47 \pm 11.06$ & $-7.69 \pm 8.98\phantom{0}$ \\
      & $D_s^+ \to \eta_{\gamma\gamma}K^+$
      & $-2.37 \pm 3.33$ & $2.32 \pm 3.23$ & $-0.02 \pm 2.32\phantom{0}$ \\
      & $D_s^+ \to K_S^0K^+$
      & $-2.042 \pm 0.536$ & $1.914 \pm 0.565$ & $-0.064 \pm 0.389\phantom{0}$ \\
    \hline
    \multirow{3}{*}{$\eta_{3\pi}K^+$}
      & $D^+ \to \eta_{3\pi}K^+$
      & $\phantom{00}11.93 \pm 13.42$ & $\phantom{0}0.03 \pm 13.25$ & $5.98 \pm 9.43$ \\
      & $D_s^+ \to \eta_{3\pi}K^+$
      & $-3.01 \pm 4.25$ & $3.40 \pm 4.14$ & $0.20 \pm 2.97$ \\
      & $D_s^+ \to K_S^0K^+$
      & $-2.084 \pm 0.453$ & $2.314 \pm 0.482$ & $0.115 \pm 0.331$ \\
  \end{tabular}
  \end{ruledtabular}
\end{table*}

\section{Branching fractions of \texorpdfstring{$D^+_{(s)} \to \eta K^+$}{D(s)+ to eta K+}}
\label{sec:branching_fraction}

The branching fractions of $D^+_{(s)} \to \eta K^+$ are measured relative to the corresponding normalization modes $D^+_{(s)} \to \eta \pi^+$, using Eq.~\ref{br_ratio_equation}.
The ratio is determined separately for the two $\eta$ reconstruction channels, $\eta\to\gamma\gamma$ and $\eta\to\pi^+\pi^-\pi^0$.
The same BDT selection trained for $D^+_{(s)}\to\eta K^+$ is applied to both the signal and normalization modes.
This common selection reduces possible systematic effects associated with all selections including BDT efficiency.
For the branching fraction measurement, additional pseudoexperiment studies are performed for the fitted signal yields.
The fitted yields show no bias and good linearity with respect to the injected signal yields, along with appropriately scaled uncertainties.

The signal yields are obtained from the invariant-mass fits summarized in Table~\ref{tab:eff_yields}.
For the branching fraction measurement, the fits are performed by integrating over the full $\cos\theta^*$ range.
The absolute branching fractions are obtained by multiplying the measured ratios by the external world-average branching fraction values $\mathcal{B}(D^+\to\eta\pi^+) = (3.75 \pm 0.09)\times10^{-3},$ and $\mathcal{B}(D_s^+\to\eta\pi^+) = (1.686 \pm 0.027)\times10^{-2}$~\cite{PDG:2026}.

\begin{table}[t]
  \centering
  \caption{
    Reconstruction efficiencies and fitted signal yields for
    $D_{(s)}^+ \to \eta h^+$, where $h = \pi, K$.
    Efficiencies are evaluated after the BDT selection trained on
    $D_{(s)}^+ \to \eta K^+$.
  }
  \label{tab:eff_yields}
  \begin{ruledtabular}
  \begin{tabular}{lccc}
    Decay mode & $\eta$ mode & $\epsilon$ $[10^{-2}]$ & $N_{\rm sig}$ \\
    \hline
    \multirow{2}{*}{$D^+ \to \eta K^+$}
      & $\gamma\gamma$       & $5.57 \pm 0.01$ & $\phantom{00}737 \pm 76\phantom{0}$ \\
      & $\pi^+\pi^-\pi^0$    & $5.65 \pm 0.01$ & $\phantom{00}439 \pm 44\phantom{0}$ \\
    \hline
    \multirow{2}{*}{$D_s^+ \to \eta K^+$}
      & $\gamma\gamma$       & $4.89 \pm 0.01$ & $\phantom{0}3781 \pm 105$ \\
      & $\pi^+\pi^-\pi^0$    & $4.95 \pm 0.01$ & $\phantom{0}1909 \pm 61\phantom{0}$ \\
    \hline
    \multirow{2}{*}{$D^+ \to \eta \pi^+$}
      & $\gamma\gamma$       & $7.64 \pm 0.01$ & $35361 \pm 301$ \\
      & $\pi^+\pi^-\pi^0$    & $7.52 \pm 0.01$ & $19966 \pm 182$ \\
    \hline
    \multirow{2}{*}{$D_s^+ \to \eta \pi^+$}
      & $\gamma\gamma$       & $6.80 \pm 0.01$ & $55606 \pm 288$ \\
      & $\pi^+\pi^-\pi^0$    & $6.71 \pm 0.01$ & $30862 \pm 197$ \\
  \end{tabular}
  \end{ruledtabular}
\end{table}

\section{Systematic uncertainties}
\label{sec:systematics} 

\subsection{\texorpdfstring{$C\!P$ Asymmetry}{CP Asymmetry}}
\label{sec:systematics_acp}

\begin{table*}[t]
  \centering
  \caption{
    Summary of systematic uncertainties in the measurements of $A_{C\!P}(D^+_{(s)} \to \eta h^+)$, where $h^+ = \pi^+, K^+$.
    All uncertainties are given in percent.
    The control mode is $D^+ \to K_S^0\pi^+$ for the $D^+_{(s)} \to \eta\pi^+$ modes and $D_s^+ \to K_S^0K^+$ for the $D^+_{(s)} \to \eta K^+$ modes.
  }
  \label{tab:total_sys_Acp_combined}
  \begin{ruledtabular}
  \begin{tabular}{lcccccccc}
    Source
    & \multicolumn{2}{c}{$D^+ \to \eta\pi^+$}
    & \multicolumn{2}{c}{$D_s^+ \to \eta\pi^+$}
    & \multicolumn{2}{c}{$D^+ \to \eta K^+$}
    & \multicolumn{2}{c}{$D_s^+ \to \eta K^+$} \\
    \cline{2-9}
    & $\eta_{\gamma\gamma}$
    & $\eta_{3\pi}$
    & $\eta_{\gamma\gamma}$
    & $\eta_{3\pi}$
    & $\eta_{\gamma\gamma}$
    & $\eta_{3\pi}$
    & $\eta_{\gamma\gamma}$
    & $\eta_{3\pi}$ \\
    \hline
    Fit modeling of signal mode
    & $0.020$ & $0.022$ & $0.019$ & $0.024$
    & $0.409$ & $0.230$ & $0.125$ & $0.117$ \\
    Fit modeling of control mode
    & $0.009$ & $0.009$ & $0.009$ & $0.009$
    & $0.020$ & $0.013$ & $0.020$ & $0.013$ \\
    Residual kinematic difference
    & $0.000$ & $0.010$ & $0.000$ & $0.009$
    & $0.089$ & $0.185$ & $0.067$ & $0.074$ \\
    Neutral kaon asymmetry
    & $0.007$ & $0.007$ & $0.007$ & $0.007$
    & $0.007$ & $0.007$ & $0.007$ & $0.007$ \\
    \hline
    Total
    & $0.023$ & $0.027$ & $0.022$ & $0.028$
    & $0.419$ & $0.295$ & $0.143$ & $0.139$ \\
  \end{tabular}
  \end{ruledtabular}
\end{table*}

The systematic uncertainty on $A_{C\!P}$ is evaluated from three main sources: fitting procedure, residual kinematic differences between signal and control modes, and the neutral-kaon-induced asymmetry in the control channels.

The fit-modeling uncertainty is estimated by varying the assumptions used in the invariant-mass fits.
In the signal-mode fits, the effects of fixed signal-shape and peaking-background parameters are evaluated using the bootstrap method~\cite{EfronTibshirani1994}.
Each bootstrap sample is formed by randomly drawing candidates with replacement from the simulated sample and is sized to match the integrated luminosity of the data.
The fixed parameters are simultaneously varied according to Gaussian distributions centered at their nominal values, with widths given by their uncertainties.
The fit is repeated for each bootstrapped sample, and the standard deviation of the resulting distribution of fitted raw asymmetries is assigned as the systematic uncertainty.
For the control modes, the fit-modeling uncertainty is evaluated by replacing the nominal signal and background PDFs with alternative forms.
The nominal Johnson $S_U$ function convolved with a Gaussian is replaced by a double-sided Crystal Ball function convolved with a Gaussian for the signal component, while a first-order polynomial is used instead of the nominal exponential function for the background.
The absolute change in the fitted raw asymmetry is assigned as the uncertainty.

As a systematic check, possible charge-dependent effects in the floated PDF parameters are studied by introducing one charge-asymmetric shape parameter at a time; no significant charge-dependent bias is observed.

An additional uncertainty in the $\eta K^+$ fits is considered for the misidentified $D^+\to\eta\pi^+$ peaking background.
In the nominal fit, the asymmetry of this component is fixed to zero to ensure fit stability.
The fit is repeated using the value estimated from simulation and its $\pm1\sigma$ variations, with the magnitude of the largest shift from the nominal result assigned as the uncertainty.

Residual kinematic differences between the signal and control modes can prevent the detection asymmetry from canceling completely in the corrected raw asymmetry.
This effect is evaluated using the momentum and polar-angle distributions of the hard hadron obtained from correctly reconstructed signal and control simulation samples.
Because the detection asymmetry depends on these variables, the control samples are reweighted in the two-dimensional distribution to match the corresponding signal distributions.
The reweighting is performed separately in the $\cos\theta^*>0$ and $\cos\theta^*<0$ regions, and the magnitude of the resulting change in raw asymmetry in the weighted samples is taken as the uncertainty.

The uncertainty associated with the neutral-kaon-induced asymmetry is evaluated by varying the detector material density in the simulation by $\pm5\%$ and taking the larger magnitude of the resulting shifts.
When combining the two $\eta$ reconstruction modes for the same parent decay, only this uncertainty is treated as fully correlated, while all other systematic uncertainties are treated as uncorrelated.

For each decay mode, the total systematic uncertainty is obtained by adding the individual contributions in quadrature.
The total systematic uncertainties are about $0.02$--$0.03\%$ and $0.1$--$0.4\%$, for the $\eta\pi^+$ and $\eta K^+$ modes, respectively, as summarized in Table~\ref{tab:total_sys_Acp_combined}. 
The larger uncertainties for the $\eta K^+$ modes arise mainly from the limited sizes of the signal samples and the treatment of the misidentified peaking background.
Across all modes, the final systematic uncertainties are small compared with the corresponding statistical uncertainties.

\subsection{Branching fractions}
\label{sec:systematics_bf}

Systematic uncertainties on the branching-fraction measurements are evaluated for the ratios $\mathcal{B}(D^+_{(s)}\to\eta K^+)/\mathcal{B}(D^+_{(s)}\to\eta\pi^+)$.
The sources considered are particle identification, the BDT selection, fit modeling, the finite size of the signal simulation samples, misreconstructed signal candidates, and the uncertainties on the world-average branching fractions of the normalization modes. 
The resulting systematic uncertainties are summarized in Table~\ref{tab:total_sys_Br_etaKp_combined}.

\begin{table}[t]
  \centering
  \caption{
    Summary of relative systematic uncertainties in the measurements of $\mathcal{B}(D^+_{(s)} \to \eta K^+)$.
    All uncertainties are given in percent. 
    The total excluding the normalization branching-fraction uncertainty is shown separately from the total including this uncertainty.
  }
  \label{tab:total_sys_Br_etaKp_combined}
  \begin{ruledtabular}
  \begin{tabular}{lcccc}
    Source
    & \multicolumn{2}{c}{$D^+ \to \eta K^+$}
    & \multicolumn{2}{c}{$D_s^+ \to \eta K^+$} \\
    \cline{2-5}
    & $\eta_{\gamma\gamma}$
    & $\eta_{3\pi}$
    & $\eta_{\gamma\gamma}$
    & $\eta_{3\pi}$ \\
    \hline
    Particle identification
    & $0.42$ & $0.43$
    & $0.41$ & $0.41$ \\
    BDT
    & $0.73$ & $0.45$
    & $0.73$ & $0.45$ \\
    Signal-mode fit model
    & $1.96$ & $1.17$
    & $1.47$ & $1.61$ \\
    Normalization-mode fit model
    & $1.87$ & $1.05$
    & $1.47$ & $1.37$ \\
    Finite signal-simulation sample size
    & $0.22$ & $0.22$
    & $0.23$ & $0.24$ \\
    Misreconstructed signal
    & $0.04$ & $0.35$
    & $0.08$ & $0.06$ \\
    Normalization branching fraction
    & $2.39$ & $2.39$
    & $1.60$ & $1.60$ \\
    \hline
    Total excluding normalization $\mathcal{B}$
    & $2.84$ & $1.74$
    & $2.26$ & $2.21$ \\
    Total
    & $3.71$ & $2.95$
    & $2.77$ & $2.73$ \\
  \end{tabular}
  \end{ruledtabular}
\end{table}

Most reconstruction-related systematic effects largely cancel in the branching-fraction ratio because the signal and normalization modes have similar final-state topologies and are selected using the same requirements.
An exception to this is the particle identification criterion applied to the hard charged hadron.
In particular, the charged-track reconstruction uncertainty is common to numerator and denominator and is therefore expected to cancel.
The cancellation of the $\eta$ reconstruction and selection efficiency has been explicitly verified by comparing the relevant kinematic distributions and by reweighting the normalization mode to match the signal mode in $\eta$ momentum and polar angle. 
The resulting efficiency changes are within the statistical uncertainties of the corresponding signal efficiencies and are therefore considered negligible.

The particle identification uncertainty is evaluated using calibration samples of $D^{*+}\to D^0(\to K^-\pi^+)\pi^+$ decays.
Data-to-simulation particle identification efficiency ratios are obtained as functions of the momentum and polar angle of the charged hadron.
The resulting correction factors for the branching-fraction ratios are consistent with unity.
Therefore, no particle identification efficiency correction is applied, and the uncertainty of the correction factor is assigned as a systematic uncertainty and treated as fully correlated between the two $\eta$ reconstruction modes.

The BDT systematic uncertainty accounts for possible imperfect cancellation of the BDT efficiency between the $\eta K^+$ signal mode and the $\eta\pi^+$ normalization mode.
Although the same BDT requirement is applied to both modes, the BDT input distributions can differ because of the different particle identification requirements for kaons and pions.
This effect is studied using the control channels $D^+\to K_S^0K^+$ and $D^+\to K_S^0\pi^+$.
The data-to-simulation BDT efficiency ratios are compared between the kaon and pion modes, and the deviation of the double ratio from unity is checked.
The double ratios are $(99.27\pm0.55)\%$ and $(99.74\pm0.45)\%$ for the $\eta_{\gamma\gamma}K^+$ and $\eta_{3\pi}K^+$ modes, respectively, where the uncertainties are statistical.
For the $\eta_{\gamma\gamma}K^+$ mode, the value deviates from unity by $0.73\%$ $(1.34\sigma)$, and this full difference is assigned as the systematic uncertainty.
In contrast, no significant deviation is observed in the $\eta_{3\pi}K^+$ mode $(0.57\sigma)$; therefore, the $0.45\%$ uncertainty on the double ratio is used instead.
Regarding the $D_s^+\to \eta K^+$ branching-fraction measurements, production-related systematic effects cancel when normalized to the $D_s^+ \to \eta \pi^+$ channel.
The remaining BDT systematic uncertainty is assumed to be dominated by the difference between the $K^+$ and $\pi^+$ final states, an assumption that is verified using simulated samples.
The uncertainties extracted from the $D^+$ control samples are therefore applied to the corresponding $D_s^+$ decay modes with the same final states.

Fit-modeling uncertainties are evaluated by repeating the branching-fraction fits with modified fit assumptions.
The effect of fixed signal-shape parameters is estimated using the bootstrap method in the same way as the $C\!P$ asymmetry measurements.
The width of the resulting signal-yield distribution is assigned as the corresponding uncertainty.
The uncertainty from the combinatorial-background model is evaluated by replacing the nominal background PDF with an alternative function, $\exp[c_0(m-m_0)+c_1(m-m_0)^2]$, where $c_0$ and $c_1$ are floated and $m_0=1.919~\mathrm{GeV}/c^2$ is the midpoint of the nominal $D^+$ and $D^+_s$ masses~\cite{PDG:2026}.
The change in the fitted signal yield is taken as the systematic uncertainty.
An additional uncertainty in the $\eta K^+$ fits is assigned for the fixed yield of the misidentified $D^+\to\eta\pi^+$ peaking background by varying this yield within its uncertainty.

The uncertainty due to the finite size of the signal simulation samples is taken from the statistical uncertainty on the reconstruction efficiencies.
A separate uncertainty is assigned for misreconstructed signal candidates.
This uncertainty is evaluated by comparing the nominal efficiency ratio with the corresponding ratio obtained using correctly reconstructed signal candidates only.

Finally, the uncertainties on the world-average normalization branching fractions, $\mathcal{B}(D^+\to\eta\pi^+)$ and $\mathcal{B}(D_s^+\to\eta\pi^+)$, are included.
These contributions are treated as fully correlated between the two $\eta$ reconstruction modes for the same parent decay.

\section{Results}
\label{sec:results}

We measure the $C\!P$ asymmetries in $D^+_{(s)}\to\eta h^+$ decays, where $h^+=\pi^+$ or $K^+$.
The measurements are performed separately for the two $\eta$ reconstruction modes, $\eta\to\gamma\gamma$ and $\eta\to\pi^+\pi^-\pi^0$.
The results are
\begin{align*}
A_{C\!P}(D^+ \to \eta_{\gamma\gamma}\pi^+) &=
(-2.06 \pm 6.55 \pm 0.23)\times 10^{-3}, \\
A_{C\!P}(D^+ \to \eta_{3\pi}\pi^+) &=
(+1.50 \pm 8.34 \pm 0.27)\times 10^{-3}, \\
A_{C\!P}(D_s^+ \to \eta_{\gamma\gamma}\pi^+) &=
(-1.30 \pm 4.80 \pm 0.22)\times 10^{-3}, \\
A_{C\!P}(D_s^+ \to \eta_{3\pi}\pi^+) &=
(-6.46 \pm 6.20 \pm 0.28)\times 10^{-3}, \\
A_{C\!P}(D^+ \to \eta_{\gamma\gamma}K^+) &=
(-8.05 \pm 8.99 \pm 0.42)\times 10^{-2}, \\
A_{C\!P}(D^+ \to \eta_{3\pi}K^+) &=
(+5.44 \pm 9.44 \pm 0.30)\times 10^{-2}, \\
A_{C\!P}(D_s^+ \to \eta_{\gamma\gamma}K^+) &=
(-0.39 \pm 2.35 \pm 0.14)\times 10^{-2}, \\
A_{C\!P}(D_s^+ \to \eta_{3\pi}K^+) &=
(-0.34 \pm 2.98 \pm 0.14)\times 10^{-2}.
\end{align*}
The first uncertainty is statistical and the second is systematic.
Combining the two $\eta$ reconstruction modes gives
\begin{align*}
A_{C\!P}(D^+ \to \eta\pi^+) &=
(-0.70 \pm 5.15 \pm 0.18)\times 10^{-3}, \\
A_{C\!P}(D_s^+ \to \eta\pi^+) &=
(-3.23 \pm 3.80 \pm 0.18)\times 10^{-3}, \\
A_{C\!P}(D^+ \to \eta K^+) &=
(-1.63 \pm 6.51 \pm 0.26)\times 10^{-2}, \\
A_{C\!P}(D_s^+ \to \eta K^+) &=
(-0.37 \pm 1.85 \pm 0.10)\times 10^{-2}.
\end{align*}

The relative branching fractions of $D^+_{(s)}\to\eta K^+$ are measured with respect to the corresponding normalization modes $D^+_{(s)}\to\eta\pi^+$.
For the individual $\eta$ reconstruction modes, the results are
\begin{align*}
\frac{\mathcal{B}(D^{+} \to \eta_{\gamma\gamma} K^+)}
     {\mathcal{B}(D^{+} \to \eta_{\gamma\gamma} \pi^+)}
&=
(2.86 \pm 0.30 \pm 0.08)\times 10^{-2}, \\
\frac{\mathcal{B}(D^{+} \to \eta_{3\pi} K^+)}
     {\mathcal{B}(D^{+} \to \eta_{3\pi} \pi^+)}
&=
(2.93 \pm 0.30 \pm 0.05)\times 10^{-2}, \\
\frac{\mathcal{B}(D_s^{+} \to \eta_{\gamma\gamma} K^+)}
     {\mathcal{B}(D_s^{+} \to \eta_{\gamma\gamma} \pi^+)}
&=
(9.46 \pm 0.27 \pm 0.21)\times 10^{-2}, \\
\frac{\mathcal{B}(D_s^{+} \to \eta_{3\pi} K^+)}
     {\mathcal{B}(D_s^{+} \to \eta_{3\pi} \pi^+)}
&=
(8.40 \pm 0.27 \pm 0.19)\times 10^{-2}.
\end{align*}
The two $D_s^+$ measurements are compatible within $2.2\sigma$, taking into account both statistical and systematic uncertainties and correlations among the systematic components.
Combining the two $\eta$ decay modes gives
\begin{align*}
\frac{\mathcal{B}(D^{+} \to \eta K^+)}
     {\mathcal{B}(D^{+} \to \eta \pi^+)}
&=
(2.90 \pm 0.21 \pm 0.05)\times 10^{-2}, \\
\frac{\mathcal{B}(D_s^{+} \to \eta K^+)}
     {\mathcal{B}(D_s^{+} \to \eta \pi^+)}
&=
(8.91 \pm 0.19 \pm 0.14)\times 10^{-2}.
\end{align*}
Using the world-average values of branching fraction for the normalization modes, the absolute branching fractions are obtained as
\begin{align*}
\mathcal{B}(D^{+} \to \eta K^+)
&=
(1.086 \pm 0.078 \pm 0.032)\times 10^{-4} \nonumber \\
&=
(1.086 \pm 0.078 \pm 0.018 \pm 0.026)\times 10^{-4}, \\
\mathcal{B}(D_s^{+} \to \eta K^+)
&=
(1.500 \pm 0.032 \pm 0.034)\times 10^{-3} \nonumber \\
&=
(1.500 \pm 0.032 \pm 0.024 \pm 0.024)\times 10^{-3}.
\end{align*}
In the second expression for each branching fraction, the last uncertainty is due to the external branching fraction of the normalization mode.
The $D^+\to\eta K^+$ absolute branching fraction is consistent with the world-average value and provides the most precise determination to date, while the $D_s^+\to\eta K^+$ branching fraction is lower than the world average by $2.8\sigma$~\cite{PDG:2026}.
Our $D_s^+\to\eta K^+$ result is compatible with the BESIII result~\cite{Bes2020_DsptoEtaKp} within $1.0\sigma$ and differs from the Belle result~\cite{Belle2021_CPAsym_DstoEtaPi} by $2.4\sigma$.
The present measurement has the smallest uncertainty among these measurements.

\section{Summary and Outlook}
\label{sec:summary_and_outlook}

In summary, we report measurements of the $C\!P$ asymmetries in $D^+_{(s)}\to\eta h^+(h=\pi,K)$ decays, as well as measurements of the branching fractions of $D^+_{(s)}\to\eta K^+$ relative to the corresponding $D^+_{(s)}\to\eta\pi^+$ normalization modes.
The analysis uses Belle~II data corresponding to an integrated luminosity of $428\,\mathrm{fb}^{-1}$.
The measured $C\!P$ asymmetries are consistent with zero, with no evidence of $C\!P$ violation, and are in agreement with previous measurements~\cite{LHCb2023_CPAsym_DstoEtaPi,LHCb2021_CPViolation_DtoHPi0Eta,BESIII:2024oth,Belle2021_CPAsym_DstoEtaPi,Belle2011_DtoEtaHp,Cleo2010_DtoEtaHp}.
The precision achieved for the pion modes is comparable to the most precise previous measurements, while the kaon-mode results constitute the most precise determinations to date.

The $A_{C\!P}$ measurements presented in this paper could become inputs to flavor-symmetry sum rules involving some $D^+_{(s)}$ decays such as $D^+\to\pi^+\eta^{(\prime)}$, $D_s^+\to K^+\eta^{(\prime)}$, and $D_s^+\to\pi^0K^+$.
By exploiting the mixing between the octet and singlet components in $\eta-\eta^\prime$~\cite{Grossman:2013lya,Iguro:2024uuw}, these results can be used to further probe the flavor structure of $C\!P$ violation in charmed meson decays.

% \input{instructions}

% \nolinenumbers

\section*{Acknowledgements}
% Policy from October 20, 2022
This work, based on data collected using the Belle II detector, which was built and commissioned prior to March 2019,
%Belle1 and data collected using the Belle detector, which was operated until June 2010,
was supported by
%Armenia
Higher Education and Science Committee of the Republic of Armenia Grant No.~23LCG-1C011;
%Australia
Australian Research Council and Research Grants
No.~DP200101792, % Jackson
No.~DP210101900, % Urquijo
No.~DP210102831, % Sevior
No.~DE220100462, % Hsu
No.~LE210100098, % Infrastructure
and
No.~LE230100085; % Infrastructure
%Austria
Austrian Federal Ministry of Education, Science and Research,
Austrian Science Fund (FWF) Grants
DOI:~10.55776/P34529,
DOI:~10.55776/J4731,
DOI:~10.55776/J4625,
DOI:~10.55776/M3153,
and
DOI:~10.55776/PAT1836324,
and
Horizon 2020 ERC Starting Grant No.~947006 ``InterLeptons'';
%Canada
Natural Sciences and Engineering Research Council of Canada, Digital Research Alliance of Canada, and Canada Foundation for Innovation;
%China
National Key R\&D Program of China under Contract No.~2024YFA1610503,
and
No.~2024YFA1610504
National Natural Science Foundation of China and Research Grants
No.~11575017,
No.~11761141009,
No.~11705209,
No.~11975076,
No.~12135005,
No.~12150004,
No.~12161141008,
No.~12405099,
No.~12475093,
No.~12675113, 
and
No.~12175041,
and Shandong Provincial Natural Science Foundation Project~ZR2022JQ02;
%Czech Republic
the Czech Science Foundation Grant No. 22-18469S,  Regional funds of EU/MEYS: OPJAK
FORTE CZ.02.01.01/00/22\_008/0004632 
and
Charles University Grant Agency project No. 246122;
%EU
European Research Council, Seventh Framework PIEF-GA-2013-622527,
Horizon 2020 ERC-Advanced Grants No.~267104 and No.~884719,
Horizon 2020 ERC-Consolidator Grant No.~819127,
Horizon 2020 Marie Sklodowska-Curie Grant Agreement No.~700525 ``NIOBE''
and
No.~101026516,
and
Horizon Europe Marie Sklodowska-Curie Staff Exchange project JENNIFER3 Grant Agreement No.~101183137 (European grants);
%France
L’Institut National de Physique Nucl\'eaire et de Physique des
Particules (IN2P3) du CNRS under Project Identification No.
CNRS-IN2P3-14-PP-033
and L’Agence Nationale de la Recherche (ANR) under Grant No. ANR-23-CE31-
0018 and ANR-25-CE31-1333 (France);
%Germany
BMFTR, DFG, HGF, MPG, and AvH Foundation (Germany);
%India
Department of Atomic Energy under Project Identification No.~RTI 4002,
Department of Science and Technology,
and
UPES SEED funding programs
No.~UPES/R\&D-SEED-INFRA/17052023/01 and
No.~UPES/R\&D-SOE/20062022/06 (India);
%Israel
Israel Science Foundation Grant No.~2476/17,
U.S.-Israel Binational Science Foundation Grant No.~2016113, and
Israel Ministry of Science Grant No.~3-16543;
%Italy
Istituto Nazionale di Fisica Nucleare and the Research Grants BELLE2,
and
the ICSC – Centro Nazionale di Ricerca in High Performance Computing, Big Data and Quantum Computing, funded by European Union – NextGenerationEU;
%Japan
Japan Society for the Promotion of Science, Grant-in-Aid for Scientific Research Grants
No.~16H03993,
No.~16H06492,
No.~16K05323,
No.~17H01133,
No.~17H05405,
No.~18K03621,
No.~18H03710,
No.~18H05226,
No.~19H00682, % Niigata
No.~20H05850,
No.~20H05858,
No.~22H00144,
No.~22K14056,
No.~22K21347,
No.~23H05433,
No.~26220706,
No.~26400255,
and
No.~26H02056,
%the National Institute of Informatics, and Science Information NETwork 5 (SINET5), 
and
the Ministry of Education, Culture, Sports, Science, and Technology (MEXT) of Japan;  
%Korea
National Research Foundation (NRF) of Korea Grants
No.~2021R1-F1A-1064008,
No.~2022R1-A2C-1003993,
No.~RS-2018-NR031074,
No.~RS-2021-NR060129,
No.~RS-2024-00354342,
No.~RS-2025-02219521,
No.~RS-2026-25471491,
No.~RS-2026-25480677,
and
No.~RS-2026-25486791,
Radiation Science Research Institute,
Foreign Large-Size Research Facility Application Supporting project,
the Global Science Experimental Data Hub Center, the Korea Institute of Science and
Technology Information (K26L1M2C3)
and
KREONET/GLORIAD;
%Malaysia
Universiti Malaya RU grant, Akademi Sains Malaysia, and Ministry of Education Malaysia;
%Mexico
% CINVESTAV-IPN, UNAM, UAS, BUAP and CONACYT are funded under
Frontiers of Science Program Contracts
No.~FOINS-296,
No.~CB-221329,
No.~CB-236394,
No.~CB-254409,
and
No.~CB-180023, and SEP-CINVESTAV Research Grant No.~237 (Mexico);
%Poland
the Polish Ministry of Science and Higher Education and the National Science Center;
%Russia
the Ministry of Science and Higher Education of the Russian Federation
and
the HSE University Basic Research Program, Moscow;
%Saudi Arabia
University of Tabuk Research Grants
No.~S-0256-1438 and No.~S-0280-1439 (Saudi Arabia);
%Slovenia
Slovenian Research Agency and Research Grants
No.~J1-50010
and
No.~P1-0135;
%Spain
Ikerbasque, Basque Foundation for Science,
State Agency for Research of the Spanish Ministry of Science and Innovation through Grant No. PID2022-136510NB-C33, Spain,
Agencia Estatal de Investigacion, Spain
Grant No.~RYC2020-029875-I
and
Generalitat Valenciana, Spain
Grant No.~CIDEGENT/2018/020;
%Swiss (Belle 1)
%Belle1 the Swiss National Science Foundation;
%Sweden
The Knut and Alice Wallenberg Foundation (Sweden), Contracts No.~2021.0174, No.~2021.0299, and No.~2023.0315;
%Taiwan
National Science and Technology Council,
and
Ministry of Education (Taiwan);
%Thailand
Thailand Center of Excellence in Physics;
%Turkey
TUBITAK ULAKBIM (Turkey);
%Ukraine
National Research Foundation of Ukraine, Project No.~2020.02/0257,
and
Ministry of Education and Science of Ukraine;
%USA
the U.S. National Science Foundation and Research Grants
No.~PHY-1913789 % Indiana CEEM
and
No.~PHY-2111604, % Luther
and the U.S. Department of Energy and Research Awards
No.~DE-AC06-76RLO1830, % PNNL
No.~DE-SC0007983, % Wayne State
No.~DE-SC0009824, % Florida
No.~DE-SC0009973, % VPI
No.~DE-SC0010007, % Duke
No.~DE-SC0010073, % South Carolina
No.~DE-SC0010118, % Carnegie Mellon
No.~DE-SC0010504, % Hawaii
No.~DE-SC0011784, % Cincinnati
No.~DE-SC0012704, % BNL
No.~DE-SC0019230, % Duke
No.~DE-SC0021616, % Mississippi
No.~DE-SC0022350, % Louisville
No.~DE-SC0023470; % South Alabama
%last group
and
%Vietnam
the Vietnam Academy of Science and Technology (VAST) under Grant
No.~DL0000.05/26-27.
Y.-J.~K. acknowledges financial support from the Yonsei Lee Youn Jae Fellowship.

% Policy from October 20, 2022
These acknowledgements are not to be interpreted as an endorsement of any statement made
by any of our institutes, funding agencies, governments, or their representatives.

We thank the SuperKEKB team for delivering high-luminosity collisions;
the KEK cryogenics group for the efficient operation of the detector solenoid magnet and IBBelle on site;
the KEK Computer Research Center for on-site computing support; the NII for SINET6 network support;
and the raw-data centers hosted by BNL, DESY, GridKa, IN2P3, INFN, 
%Belle1 PNNL/EMSL, 
and the University of Victoria.

\section*{Data Availability}
The Belle II data underlying this article are not publicly available due to legal restrictions.
The data may be made available by the authors upon reasonable request, subject to applicable restrictions.

\bibliography{references}

@article{Buccella1995,
  author = "Buccella, F. and Lusignoli, Maurizio and Miele, G. and Pugliese, A. and Santorelli, Pietro",
  title = {Nonleptonic weak decays of charmed mesons},
  journal = {Phys. Rev. D},
  volume = {51},
  year = {1995},
  pages = {3478--3487},
  doi = {10.1103/PhysRevD.51.3478},
  archivePrefix = "arXiv",
  eprint = {hep-ph/9411286}
}

@article{Li:2012cfa,
  author       = {Li, Hsiang-nan and Lu, Cai-Dian and Yu, Fu-Sheng},
  title        = {Branching ratios and direct {$C\!P$} asymmetries in {$D\to PP$} decays},
  journal      = {Phys. Rev. D},
  volume       = {86},
  pages        = {036012},
  year         = {2012},
  doi          = {10.1103/PhysRevD.86.036012},
  archivePrefix= {arXiv},
  eprint       = {1205.3120},
  primaryClass = {hep-ph}
}

@article{Cheng:2012xb,
    author = "Cheng, Hai-Yang and Chiang, Cheng-Wei",
    title = {{SU(3)} symmetry breaking and {$C\!P$} violation in {$D\to P P$} decays},
    journal      = {Phys. Rev. D},
    volume       = {86},
    pages        = {014014},
    year         = {2012},
    doi          = {10.1103/PhysRevD.86.014014},
    archivePrefix= {arXiv},
    eprint       = {1205.0580},
    primaryClass = {hep-ph}
}

@article{Wolfen:1983,
  author       = {Lincoln Wolfenstein},
  title        = {Parametrization of the {Kobayashi-Maskawa} matrix},
  journal      = {Phys. Rev. Lett.},
  volume       = {51},
  pages        = {1945},
  year         = {1983},
  doi          = {10.1103/PhysRevLett.51.1945},
}

@article{LHCb2023_CPAsym_DstoEtaPi,
    author = "Aaij, Roel and others",
    collaboration = "LHCb Collaboration",
    title = {Measurement of {$C\!P$} asymmetries in {$D^+_{(s)} \to \eta \pi^+$} and {$D^+_{(s)} \to \eta^{\prime} \pi^+$} decays},
    journal = "JHEP",
    volume  = {04},
    number = {2023},
    year = {2023},
    pages = {081},
    doi = "10.1007/JHEP04(2023)081",
    archivePrefix = "arXiv",
    eprint = "2204.12228",
    primaryClass = "hep-ex"
}

@article{LHCb2021_CPViolation_DtoHPi0Eta,
  author = "Aaij, Roel and others",
  collaboration = "LHCb Collaboration",
  title        = {Search for {$C\!P$} violation in {$D^+_{(s)} \to h^+ \pi^0$} and {$D^+_{(s)} \to h^+ \eta$} decays},
  journal      = {JHEP},
  volume       = {06},
  pages        = {019},
  year         = {2021},
  number       = {2021},
  doi          = {10.1007/JHEP06(2021)019},
  eprint       = {2103.11058},
  archivePrefix= {arXiv},
  primaryClass = {hep-ex}
}

@article{BESIII:2024oth,
    author = "Ablikim, Medina and others",
    collaboration = "BESIII Collaboration",
    title = {Measurement of absolute branching fractions of {$D_s^{+}$} hadronic decays},
    eprint = "2403.19256",
    archivePrefix = "arXiv",
    primaryClass = "hep-ex",
    doi = "10.1007/JHEP05(2024)335",
    journal = "JHEP",
    volume = "05",
    pages = "335",
    year = "2024",
    number = "2024"
}

@article{Belle2021_CPAsym_DstoEtaPi,
  author = "Guan, Y. and others",
  collaboration = "Belle Collaboration",
  title        = {Measurement of branching fractions and {$C\!P$} asymmetries for {$D^+_{s} \to K^+ (\eta,\pi^0)$} and {$D^+_{s} \to \pi^+ (\eta,\pi^0)$} decays at {Belle}},
  journal      = {Phys. Rev. D},
  volume       = {103},
  pages        = {112005},
  year         = {2021},
  doi          = {10.1103/PhysRevD.103.112005},
  eprint       = {2103.09969},
  archivePrefix= {arXiv},
  primaryClass = {hep-ex}
}

@article{Belle2011_DtoEtaHp,
  author = "Won, E. and others",
  collaboration = "Belle Collaboration",
  title        = {Observation of {$D^+ \to K^+ \eta^{(\prime)}$} and search for {$C\!P$} violation in {$D^+ \to \pi^+ \eta^{(\prime)}$} decays},
  journal      = {Phys. Rev. Lett.},
  volume       = {107},
  pages        = {221801},
  year         = {2011},
  doi          = {10.1103/PhysRevLett.107.221801},
  eprint       = {1107.0553},
  archivePrefix= {arXiv},
  primaryClass = {hep-ex}
}

@article{Cleo2010_DtoEtaHp,
  author = "Mendez, H. and others",
  collaboration = "CLEO Collaboration",
  title        = {Measurement of {$D$} meson decays to two pseudoscalar mesons},
  journal      = {Phys. Rev. D},
  volume       = {81},
  pages        = {052013},
  year         = {2010},
  doi          = {10.1103/PhysRevD.81.052013},
  eprint       = {0906.3198},
  archivePrefix= {arXiv},
  primaryClass = {hep-ex}
}

@article{Bes2018_DptoEtaKp,
  author = "Ablikim, Medina and others",
  collaboration = "BESIII Collaboration",
  title        = {Measurements of absolute branching fractions for {$D$} mesons decays into two pseudoscalar mesons},
  journal      = {Phys. Rev. D},
  volume       = {97},
  pages        = {072004},
  year         = {2018},
  doi          = {10.1103/PhysRevD.97.072004},
  eprint       = {1802.03119},
  archivePrefix= {arXiv},
  primaryClass = {hep-ex}
}

@article{Bes2025_DptoEtaKp,
  author = "Ablikim, Medina and others",
  collaboration = "BESIII Collaboration",
  title        = {Measurement of the absolute branching fractions of the doubly {Cabibbo-suppressed} decays {$D^+\to K^+ \pi^0$}, {$D^+ \to K^+ \eta$} and {$D^+ \to K^+ \eta^\prime$}},
  journal      = {JHEP},
  volume       = {10},
  pages        = {178},
  year         = {2025},
  number       = {2025},
  doi          = {10.1007/JHEP10(2025)178},
  eprint       = {2506.15533},
  archivePrefix= {arXiv},
  primaryClass = {hep-ex}
}

@article{Bes2020_DsptoEtaKp,
  author = "Ablikim, Medina and others",
  collaboration = "BESIII Collaboration",
  title        = {Precise measurements of branching fractions for {$D^+_s$} meson decays to two pseudoscalar mesons},
  journal      = {JHEP},
  volume       = {08},
  pages        = {146},
  year         = {2020},
  number       = {2020},
  doi          = {10.1007/JHEP08(2020)146},
  eprint       = {2005.05072},
  archivePrefix= {arXiv},
  primaryClass = {hep-ex}
}

@article{LHCb:2019hro,
    author = "Aaij, Roel and others",
    collaboration = {LHCb Collaboration},
    title = {Observation of {$C\!P$} violation in charm decays},
    eprint = "1903.08726",
    archivePrefix = "arXiv",
    primaryClass = "hep-ex",
    doi = "10.1103/PhysRevLett.122.211803",
    journal = "Phys. Rev. Lett.",
    volume = "122",
    number = "21",
    pages = "211803",
    year = "2019"
}

@article{Lyons1988,
  author  = {L. Lyons and D. Gibaut and P. Clifford},
  title   = {How to combine correlated estimates of a single physical quantity},
  journal = {Nucl. Instrum. Meth. A},
  volume  = {270},
  pages   = {110--117},
  year    = {1988},
  doi     = {10.1016/0168-9002(88)90018-6},
}

@article{Valassi2003,
  author  = {A. Valassi},
  title   = {Combining correlated measurements of several different physical quantities},
  journal = {Nucl. Instrum. Meth. A},
  volume  = {500},
  pages   = {391--405},
  year    = {2003},
  doi     = {10.1016/S0168-9002(03)00329-2},
}

@article{Berends:1973fd,
    author = "Berends, Frits A. and Gaemers, K. J. F. and Gastmans, R.",
    title = {{$\alpha^3$}-contribution to the angular asymmetry in {$e^+ e^- \to \mu^+ \mu^-$}},
    doi = "10.1016/0550-3213(73)90153-3",
    journal = "Nucl. Phys. B",
    volume = "63",
    pages = "381--397",
    year = "1973"
}

@article{Brown:1973ji,
    author = "Brown, R. W. and Mikaelian, K. O. and Cung, V. K. and Paschos, E. A.",
    title = {Electromagnetic background in the search for neutral weak currents via {$e^+ e^- \to \mu^+ \mu^-$}},
    doi = "10.1016/0370-2693(73)90384-5",
    journal = "Phys. Lett. B",
    volume = "43",
    pages = "403--407",
    year = "1973"
}

@article{Cashmore:1985vp,
    author = "Cashmore, R. J. and Hawkes, C. M. and Lynn, B. W. and Stuart, R. G.",
    title = {The forward-backward asymmetry in {$e^+ e^- \to \mu^+ \mu^-$}},
    reportNumber = "OXFORD-NP-45-85",
    doi = "10.1007/BF01560685",
    journal = "Z. Phys. C",
    volume = "30",
    pages = "125--134",
    year = "1986"
}

@article{LHCb:2014kcb,
    author = "Aaij, Roel and others",
    collaboration = {LHCb Collaboration},
    title = "{Measurement of {$C\!P$} asymmetry in $D^0 \rightarrow K^- K^+$ and $D^0 \rightarrow \pi^- \pi^+$ decays}",
    eprint = "1405.2797",
    archivePrefix = "arXiv",
    primaryClass = "hep-ex",
    reportNumber = "CERN-PH-EP-2014-082, LHCB-PAPER-2014-013",
    doi = "10.1007/JHEP07(2014)041",
    journal = "JHEP",
    volume = "07",
    pages = "041",
    year = "2014",
    number = "2014"
}

@article{Golden:1989qx,
    author = "Golden, Mitchell and Grinstein, Benjamin",
    title = {Enhanced {$C\!P$} violations in hadronic charm decays},
    reportNumber = "FERMILAB-PUB-89-048-T",
    doi = "10.1016/0370-2693(89)90353-5",
    journal = "Phys. Lett. B",
    volume = "222",
    pages = "501--506",
    year = "1989"
}

@inproceedings{Chen:2016btl,
    author = "Chen, Tianqi and Guestrin, Carlos",
    title = {{XGBoost}: A Scalable Tree Boosting System},
    booktitle = {Proceedings of the 22nd ACM SIGKDD international conference on knowledge discovery and data mining},
    series = "KDD '16",
    publisher = {Association for Computing Machinery},
    pages     = {785--794},
    year      = {2016},
    doi = "10.1145/2939672.2939785",
    eprint = "1603.02754",
    archivePrefix = "arXiv",
    primaryClass = "cs.LG",    
}

@article{Cheema:2024iek,
    author = "Cheema, Priyanka",
    title = {Suppressing beam background and fake photons at {Belle II} using machine learning},
    doi = "10.1051/epjconf/202429509035",
    journal = "EPJ Web Conf.",
    volume = "295",
    pages = "09035",
    year = "2024"
}

@article{PDG:2026,
    author = {Takahashi, F. and others},
    collaboration = {Particle Data Group},
    title = {Review of Particle Physics},
    journal = {Int. J. Mod. Phys. A},
    volume = {41},
    pages = {2630011},
    year = {2026}, 
    url = {https://pdg.lbl.gov/},
    doi = {10.1142/S0217751X26300115}
}

@article{Belle-II:2025tpe,
    author = "Adachi, I. and others",
    collaboration = "Belle II Collaboration",
    title = {Charged-hadron identification at {Belle II}},
    eprint = "2506.04355",
    archivePrefix = "arXiv",
    primaryClass = "hep-ex",
    reportNumber = "Belle II Preprint 2025-016, KEK Preprint 2025-15",
    doi = "10.1140/epjc/s10052-025-14627-7",
    journal = "Eur. Phys. J. C",
    volume = "85",
    number = "11",
    pages = "1237",
    year = "2025"
}

@article{Cranmer:2000du,
    author = "Cranmer, Kyle S.",
    title = "{Kernel estimation in high-energy physics}",
    eprint = "hep-ex/0011057",
    archivePrefix = "arXiv",
    doi = "10.1016/S0010-4655(00)00243-5",
    journal = "Comput. Phys. Commun.",
    volume = "136",
    pages = "198--207",
    year = "2001"
}

@article{Belle:1999bhb,
    author = "Ikeda, H. and others",
    collaboration = "Belle Collaboration",
    title = "{A detailed test of the CsI(Tl) calorimeter for BELLE with photon beams of energy between 20MeV and 5.4GeV}",
    reportNumber = "KEK-PREPRINT-99-70, BELLE-PREPRINT-99-3, NWU-HEP-99-02, TIT-HPE-99-10",
    doi = "10.1016/S0168-9002(99)00992-4",
    journal = "Nucl. Instrum. Meth. A",
    volume = "441",
    pages = "401--426",
    year = "2000"
}

@article{Johnson:1949zj,
    author = "Johnson, N. L.",
    title = "{Systems of frequency curves generated by methods of translation}",
    doi = "10.1093/biomet/36.1-2.149",
    journal = "Biometrika",
    volume = "36",
    pages = "149--176",
    year = "1949"
}

@book{EfronTibshirani1994,
author    = {Efron, Bradley and Tibshirani, Robert J.},
title     = {An Introduction to the Bootstrap},
publisher = {Chapman and Hall/CRC},
address   = {New York},
year      = {1994},
doi       = {10.1201/9780429246593}
}

@article{Cheng:2019ggx,
    author = "Cheng, Hai-Yang and Chiang, Cheng-Wei",
    title = {Revisiting {$C\!P$} violation in {$D\to PP$} and {$VP$} decays},
    eprint = "1909.03063",
    archivePrefix = "arXiv",
    primaryClass = "hep-ph",
    doi = "10.1103/PhysRevD.100.093002",
    journal = "Phys. Rev. D",
    volume = "100",
    number = "9",
    pages = "093002",
    year = "2019"
}

@article{Belle-II:2025wsy,
    author = "Adachi, Ichiro and others",
    collaboration = "Belle II Collaboration",
    title = "{Measurement of the $C\!P$ asymmetry in $D^+\to\pi^+ \pi^0$ decays at Belle II}",
    eprint = "2506.07879",
    archivePrefix = "arXiv",
    primaryClass = "hep-ex",
    reportNumber = "Belle II Preprint 2025-012, KEK Preprint 2025-10",
    doi = "10.1103/3kqk-ldm4",
    journal = "Phys. Rev. D",
    volume = "112",
    number = "3",
    pages = "L031101",
    year = "2025"
}

@article{Iguro:2024uuw,
    author = {Iguro, Syuhei and Nierste, Ulrich and Overduin, Emil and Sch{\"u}{\ss}ler, Maurice},
    title = {{$\mathrm{SU(3)_F}$} sum rules for {$C\!P$} asymmetries of {$D_{(s)}$} decays},
    eprint = "2408.03227",
    archivePrefix = "arXiv",
    primaryClass = "hep-ph",
    doi = "10.1103/PhysRevD.111.035023",
    journal = "Phys. Rev. D",
    volume = "111",
    number = "3",
    pages = "035023",
    year = "2025"
}

@article{Grossman:2013lya,
    author = "Grossman, Yuval and Ligeti, Zoltan and Robinson, Dean J.",
    title = {More flavor {SU(3)} tests for new physics in {CP} violating {$B$} decays},
    eprint = "1308.4143",
    archivePrefix = "arXiv",
    primaryClass = "hep-ph",
    doi = "10.1007/JHEP01(2014)066",
    journal = "JHEP",
    volume = "01",
    pages = "066",
    year = "2014",
    number = "2014"
}

@article{Kobayashi:1973,
      author         = "Kobayashi, Makoto and Maskawa, Toshihide",
      title          = "{\CP violation in the renormalizable theory of weak
                        interaction}",
      journal        = "Prog. Theor. Phys.",
      volume         = "49",
      year           = "1973",
      pages          = "652",
      doi            = "10.1143/PTP.49.652",
      reportNumber   = "KUNS-242",
      SLACcitation   = "%%CITATION = PTPKA,49,652;%%"
}

@misc{Abe:2010,
      author = "Abe, T. and others",
      collaboration  = "Belle II Collaboration",
      title          = "{Belle II technical design report}",
      year           = "2010",
      eprint         = "1011.0352",
      archivePrefix  = "arXiv",
      primaryClass   = "physics.ins-det",
      reportNumber   = "KEK-REPORT-2010-1",
      SLACcitation   = "%%CITATION = ARXIV:1011.0352;%%",
}

@article{Akai:2018,
      author         = "Akai, Kazunori and Furukawa, Kazuro and Koiso, Haruyo",
      title          = "{SuperKEKB collider}",
      journal        = "Nucl. Instrum. Meth. A",
      volume         = "907",
      year           = "2018",
      pages          = "188",
      doi            = "10.1016/j.nima.2018.08.017",
      eprint         = "1809.01958",
      archivePrefix  = "arXiv",
      primaryClass   = "physics.acc-ph",
      SLACcitation   = "%%CITATION = ARXIV:1809.01958;%%"
}

@article{Belle-IISVD:2022,
    author = "Adamczyk, K. and others",
    collaboration = "Belle II SVD Group",
    title = "{The design, construction, operation and performance of the Belle~II silicon vertex detector}",
    eprint = "2201.09824",
    archivePrefix = "arXiv",
    primaryClass = "physics.ins-det",
    doi = "10.1088/1748-0221/17/11/P11042",
    journal = {{JINST}},
    volume = "17",
    number = "11",
    pages = "P11042",
    year = "2022"
}

@article{Lange:2001uf,
      author         = "Lange, D. J.",
      title          = "{The EvtGen particle decay simulation package}",
      booktitle      = "{Proceedings, 7th International Conference on \B physics
                        at hadron machines (BEAUTY 2000): Maagan, Israel,
                        September 13-18, 2000}",
      journal        = "Nucl. Instrum. Meth. A",
      volume         = "462",
      year           = "2001",
      pages          = "152",
      doi            = "10.1016/S0168-9002(01)00089-4",
      SLACcitation   = "%%CITATION = NUIMA,A462,152;%%"
}

@article{Sjostrand:2014zea,
      author         = {Sj\"{o}strand, Torbj\"{o}rn and Ask, Stefan and Christiansen,
                        Jesper R. and Corke, Richard and Desai, Nishita and Ilten,
                        Philip and Mrenna, Stephen and Prestel, Stefan and
                        Rasmussen, Christine O. and Skands, Peter Z.},
      title          = "{An Introduction to PYTHIA 8.2}",
      journal        = "Comput. Phys. Commun.",
      volume         = "191",
      year           = "2015",
      pages          = "159-177",
      doi            = "10.1016/j.cpc.2015.01.024",
      eprint         = "1410.3012",
      archivePrefix  = "arXiv",
      primaryClass   = "hep-ph",
      reportNumber   = "LU-TP-14-36, MCNET-14-22, CERN-PH-TH-2014-190,
                        FERMILAB-PUB-14-316-CD, DESY-14-178, SLAC-PUB-16122",
      SLACcitation   = "%%CITATION = ARXIV:1410.3012;%%"
}

@article{Jadach:1999vf,
      author         = "Jadach, S. and Ward, B. F. L. and W\c{a}s, Z.",
      title          = "{The precision Monte Carlo event generator KK for two-fermion
                        final states in $e^+e^-$ collisions}",
      journal        = "Comput. Phys. Commun.",
      volume         = "130",
      year           = "2000",
      pages          = "260",
      doi            = "10.1016/S0010-4655(00)00048-5",
      eprint         = "hep-ph/9912214",
      archivePrefix  = "arXiv",
      primaryClass   = "hep-ph",
      reportNumber   = "DESY-99-106, CERN-TH-99-235, UTHEP-99-08-01",
      SLACcitation   = "%%CITATION = HEP-PH/9912214;%%"
}

@article{Jadach:1990mz,
      author         = "Jadach, S. and Kuhn, J. H. and W\c{a}s, Z.",
      title          = "{TAUOLA: A library of Monte Carlo programs to simulate
                        decays of polarized tau leptons}",
      journal        = "Comput. Phys. Commun.",
      volume         = "64",
      year           = "1990",
      pages          = "275",
      doi            = "10.1016/0010-4655(91)90038-M",
      reportNumber   = "CERN-TH-5856-90",
      SLACcitation   = "%%CITATION = CPHCB,64,275;%%"
}

@article{Barberio:1990ms,
      author         = "Barberio, Elisabetta and van Eijk, Bob and W\c{a}s, Zbigniew",
      title          = "{PHOTOS: A universal Monte Carlo for QED radiative
                        corrections in decays}",
      journal        = "Comput. Phys. Commun.",
      volume         = "66",
      year           = "1991",
      pages          = "115",
      doi            = "10.1016/0010-4655(91)90012-A",
      reportNumber   = "CERN-TH-5857-90",
      SLACcitation   = "%%CITATION = CPHCB,66,115;%%"
}

@article{Agostinelli:2002hh,
      author         = "Agostinelli, S. and others",
      title          = "{GEANT4: A simulation toolkit}",
      collaboration  = "GEANT4 Collaboration",
      journal        = "Nucl. Instrum. Meth. A",
      volume         = "506",
      pages          = "250-303",
      doi            = "10.1016/S0168-9002(03)01368-8",
      year           = "2003",
      reportNumber   = "SLAC-PUB-9350, FERMILAB-PUB-03-339",
      SLACcitation   = "%%CITATION = NUIMA,A506,250;%%",
}

@article{Kuhr:2018lps,
      author         = "Kuhr, T. and Pulvermacher, C. and Ritter, M. and Hauth,
                        T. and Braun, N.",
      title          = {The {Belle II} Core Software},
      collaboration  = "Belle II Framework Software Group",
      journal        = "Comput. Softw. Big Sci.",
      volume         = "3",
      year           = "2019",
      number         = "1",
      pages          = "1",
      doi            = "10.1007/s41781-018-0017-9",
      eprint         = "1809.04299",
      archivePrefix  = "arXiv",
      primaryClass   = "physics.comp-ph",
      SLACcitation   = "%%CITATION = ARXIV:1809.04299;%%"
}

@misc{basf2-zenodo,
    author = "{Belle II Collaboration}",
    title = {{Belle II} Analysis Software Framework (basf2)},
    howpublished = "\url{https://doi.org/10.5281/zenodo.5574115}"
}

@article{Bertacchi:2020eez,
    author = "Bertacchi, Valerio and others",
    collaboration = "Belle II Tracking Group",
    title = "{Track finding at Belle II}",
    eprint = "2003.12466",
    archivePrefix = "arXiv",
    primaryClass = "physics.ins-det",
    doi = "10.1016/j.cpc.2020.107610",
    journal = "Comput. Phys. Commun.",
    volume = "259",
    pages = "107610",
    year = "2021"
}

@article{Krohn:2019dlq,
    author = "Krohn, J.-F. and others",
    collaboration = "Belle II Analysis Software Group",
    title = "{Global decay chain vertex fitting at Belle II}",
    eprint = "1901.11198",
    archivePrefix = "arXiv",
    primaryClass = "hep-ex",
    doi = "10.1016/j.nima.2020.164269",
    journal = "Nucl. Instrum. Meth. A",
    volume = "976",
    pages = "164269",
    year = "2020"
}

@phdthesis{Gaiser:Phd,
    author = "Gaiser, John",
    title = "{Charmonium spectroscopy from radiative decays of the $J/\psi$ and $\psi^\prime$}",
    reportNumber = "SLAC-R-255",
    school = "Stanford University",
    year = "1982"
}

@phdthesis{Skwarnicki:1986xj,
    author = "Skwarnicki, Tomasz",
    title = "{A study of the radiative CASCADE transitions between the Upsilon-Prime and Upsilon resonances}",
    reportNumber = "DESY-F31-86-02, DESY-F-31-86-02",
    school = "Cracow, INP",
    year = "1986"
}

@article{Pivk:2004ty,
      author         = "Pivk, Muriel and Le Diberder, Francois R.",
      title          = "{sPlot: A statistical tool to unfold data distributions}",
      journal        = "Nucl. Instrum. Meth. A",
      volume         = "555",
      pages          = "356-369",
      doi            = "10.1016/j.nima.2005.08.106",
      year           = "2005",
      eprint         = "physics/0402083",
      archivePrefix  = "arXiv",
      primaryClass   = "physics.data-an",
      reportNumber   = "LAL-04-07",
      SLACcitation   = "%%CITATION = PHYSICS/0402083;%%",
}

@article{Kotchetkov:2018qzw,
    author = "Kotchetkov, Dmitri and others",
    title = "{Front-end electronic readout system for the Belle II imaging Time-Of-Propagation detector}",
    eprint = "1804.10782",
    archivePrefix = "arXiv",
    primaryClass = "physics.ins-det",
    doi = "10.1016/j.nima.2019.162342",
    journal = "Nucl. Instrum. Meth. A",
    volume = "941",
    pages = "162342",
    year = "2019"
}

@article{Pais:1955sm,
    author = "Pais, A. and Piccioni, O.",
    title = {Note on the Decay and Absorption of the {$\Theta^0$}},
    doi = "10.1103/PhysRev.100.1487",
    journal = "Phys. Rev.",
    volume = "100",
    pages = "1487--1489",
    year = "1955"
}

@article{Ko:2010mk,
    author = "Ko, B. R. and Won, E. and Golob, B. and Pakhlov, P.",
    title = {Effect of nuclear interactions of neutral kaons on {$C\!P$} asymmetry measurements},
    eprint = "1006.1938",
    archivePrefix = "arXiv",
    primaryClass = "hep-ex",
    doi = "10.1103/PhysRevD.84.111501",
    journal = "Phys. Rev. D",
    volume = "84",
    pages = "111501",
    year = "2011"
}

\end{document}